\documentclass[twocolumn,showpacs,preprintnumbers,prd,nofootinbib]{revtex4}
\usepackage{epsfig,amsmath,amssymb}

\pdfoutput=1 

\newcommand{\beq}{\begin{equation}}
\newcommand{\eeq}{\end{equation}}
\newcommand{\bea}{\begin{eqnarray}}
\newcommand{\eea}{\end{eqnarray}}
\newcommand{\Fig}[1]{Fig.\,\ref{#1}}

\newcommand{\Eq}[1]{Eq.\,(\ref{#1})}

\newcommand{\Eqsand}[2]{Eqs.\,(\ref{#1}) and (\ref{#2})}

\newcommand{\Tab}[1]{Tab.\,\ref{#1}}

\newcommand{\Tabsand}[2]{Tabs.\,\ref{#1} and \ref{#2}}
\newcommand{\Sec}[1]{Sec.\,\ref{#1}}

\newcommand{\Secsand}[2]{Secs.\,\ref{#1} and \ref{#2}}

\newcommand{\App}[1]{App.\,\ref{#1}}
\newcommand{\Appsand}[2]{Apps.\,\ref{#1} and \ref{#2}}
\newcommand{\f}{\frac}
\newcommand{\non}{\nonumber}

\newcommand{\as}{\alpha_s}

\newcommand{\sw}{s_{\scriptscriptstyle W}}
\newcommand{\cw}{c_{\scriptscriptstyle W}}
\newcommand{\sws}{s^2_{\scriptscriptstyle W}}

\newcommand{\MW}{M_{\scriptscriptstyle W}}
\newcommand{\MZ}{M_{\scriptscriptstyle Z}}

\newcommand{\MHpm}{M_H^\pm}
\newcommand{\MHpms}{M_H^{\pm 2}}
\newcommand{\MHpmq}{M_H^{\pm 4}}
\newcommand{\MTp}{M_{T_+}}
\newcommand{\MTps}{M^2_{T_+}}
\newcommand{\MTpq}{M^4_{T_+}}
\newcommand{\MTpt}{M^6_{T_+}}
\newcommand{\mtk}{m_{t (k)}}
\newcommand{\mtks}{m_{t (k)}^2}

\newcommand{\MWk}{M_{{\scriptscriptstyle W} (k)}}
\newcommand{\MWks}{M_{{\scriptscriptstyle W} (k)}^2}

\newcommand{\mk}{m_{(k)}}
\newcommand{\mks}{m_{(k)}^2}

\newcommand{\mc}{m_c}
\newcommand{\mb}{m_b}
\newcommand{\mt}{m_t}

\newcommand{\muw}{\mu_{\scriptscriptstyle W}}

\newcommand{\mub}{\mu_b}
\newcommand{\mut}{\mu_t}

\newcommand{\mstone}{m_{\tilde{t}_1}}

\newcommand{\mcha}{M_{\tilde \chi_1}^\pm}
\newcommand{\msq}{m_{\tilde{q}}}
\newcommand{\msl}{m_{\tilde{l}}}
\newcommand{\mgl}{M_{\tilde g}}
\newcommand{\mh}{M_h^0}

\newcommand{\GF}{G_F}

\newcommand{\MeV}{{\rm MeV}}
\newcommand{\GeV}{{\rm GeV}}
\newcommand{\TeV}{{\rm TeV}}
\newcommand{\MSbar}{\overline{\rm MS}}
\newcommand{\re}{{\rm Re}}

\def\unit{\leavevmode\hbox{\small1\kern-3.6pt\normalsize1}}
\newcommand{\eps}{\epsilon}
\newcommand{\gam}{\gamma}
\newcommand{\sL}{{\scalebox{0.6}{$L$}}}

\newcommand{\Kpnn}{K^+ \to \pi^+ \nu \bar{\nu}}
\newcommand{\KLnn}{K_L \to \pi^0 \nu \bar{\nu}}
\newcommand{\Knns}{K \to \pi \nu \bar{\nu}}
\newcommand{\KLmm}{K_L \to \mu^+ \mu^-}
\newcommand{\KLpill}{K_L \to \pi^0 l^+ l^-}

\newcommand{\BXsga}{\bar{B} \to X_s \gamma}

\newcommand{\BXsll}{\bar{B} \to X_s l^+ l^-}
\newcommand{\BXdsll}{\bar{B} \to X_{d,s} l^+ l^-}
\newcommand{\BXdnn}{\bar{B} \to X_d \nu \bar{\nu}}
\newcommand{\BXsnn}{\bar{B} \to X_s \nu \bar{\nu}}
\newcommand{\BXdsnn}{\bar{B} \to X_{d,s} \nu \bar{\nu}}
\newcommand{\Bdmm}{B_d \to \mu^+ \mu^-}
\newcommand{\Bsmm}{B_s \to \mu^+ \mu^-}
\newcommand{\Bdsmm}{B_{d,s} \to \mu^+ \mu^-}
\newcommand{\BRKp}{{\cal B} (\Kpnn)}
\newcommand{\BRKL}{{\cal B} (\KLnn)}
\newcommand{\BRKm}{{\cal B} (\KLmm)_{\rm SD}}
\newcommand{\BRXd}{{\cal B} (\BXdnn)}
\newcommand{\BRXs}{{\cal B} (\BXsnn)}
\newcommand{\BRBd}{{\cal B} (\Bdmm)}
\newcommand{\BRBs}{{\cal B} (\Bsmm)}
\newcommand{\BRga}{{\cal B} (\BXsga)}
\newcommand{\BRll}{{\cal B} (\BXsll)}

\newcommand{\BXuen}{\bar{B} \to X_u e \bar{\nu}}
\newcommand{\BXcen}{\bar{B} \to X_c e \bar{\nu}}
\newcommand{\btouen}{b \to u e \bar{\nu}}

\newcommand{\BRXc}{{\cal B} (\bar{B} \to X_c e \bar{\nu})}
\newcommand{\AFBBKll}{A_{\rm FB} (\bar{B} \to K^\ast l^+ l^-)} 
\newcommand{\epseps}{\epsilon^\prime/\epsilon}
\newcommand{\stodnn}{s \to d \nu \bar{\nu}}
\newcommand{\btosgamma}{b \to s \gamma}
\newcommand{\btodsll}{b \to d (s) l^+ l^-}
\newcommand{\btosll}{b \to s l^+ l^-}
\newcommand{\Ztobb}{Z \to b \bar{b}}

\newcommand{\Ztodjdi}{Z \to d_j \bar{d}_i}
\newcommand{\Zbb}{Z b \bar{b}}
\newcommand{\Zdjdi}{Z d_j \bar{d_i}}

\newcommand{\eeff}{e^+ e^- \to f \bar{f}}
\newcommand{\didjnn}{d_i \to d_j \nu \bar{\nu}} 
\newcommand{\didjll}{d_i \to d_j l^+ l^-} 
\newcommand{\X}{X}
\newcommand{\Y}{Y}
\newcommand{\Z}{Z}
\newcommand{\Bnn}{B^{\nu \bar{\nu}}}
\newcommand{\Bll}{B^{l^+ l^-}}

\newcommand{\BllSM}{B^{l^+ l^-}_{\rm SM}}
\newcommand{\C}{C}
\newcommand{\D}{D}

\newcommand{\Rb}{R_b^0}
\newcommand{\Ab}{{\cal A}_b}
\newcommand{\Al}{{\cal A}_e}
\newcommand{\AFB}{A_{\rm FB}^{0, b}}
\newcommand{\ee}{e^+ e^-}
\newcommand{\eetoU}{e^+ e^- \to \Upsilon (4S)}
\newcommand{\CP}{C\hspace{-0.25mm}P}
\newcommand{\Daehad}{\Delta \alpha_{\rm had}^{(5)} (\MZ)}
\newcommand{\mysigma}{\hspace{0.4mm} \sigma}

\newcommand{\etal}{{\it et al}.}

\begin{document}

\allowdisplaybreaks

\preprint{ZU-TH 6/07; CLNS 07/2001}

\title{
\boldmath
Determining the Sign of the $Z$-Penguin Amplitude
\unboldmath
}

\author{Ulrich~Haisch$^1$ and Andreas~Weiler$^2$} 

\affiliation{
$^1\!\!\!$ Institut f\"ur Theoretische Physik, Universit\"at Z\"urich,
CH-8057 Z\"urich, Switzerland \\
$^2\!\!\!$ Institute for High Energy Phenomenology 
Newman Laboratory of Elementary Particle Physics, 
Cornell University, Ithaca, NY 14853, U.S.A.
}

\date{\today}

\begin{abstract}
\noindent
We point out that the precision measurements of the pseudo observables 
$\Rb$, $\Ab$, and $\AFB$ performed at LEP and SLC suggest that in
models with minimal-flavor-violation the sign of the $Z$-penguin
amplitude is identical to the one present in the standard model. We
determine the allowed range for the non-standard contribution to the
Inami-Lim function $\C$ and show by analyzing possible scenarios
with positive and negative interference of standard model and new
physics contributions, that the derived bound holds in each given
case. Finally, we derive lower and upper limits for the branching
ratios of $\Kpnn$, $\KLnn$, $\KLmm$, $\BXdsnn$, and $\Bdsmm$ within
constrained minimal-flavor-violation making use of the wealth of
available data collected at the $Z$-pole.        
\end{abstract}


\maketitle

\section{Introduction}
\label{sec:introduction}

The effects of new heavy degrees of freedom appearing in extensions of
the standard model (SM) can be accounted for at low energies in terms
of effective operators. The unprecedented accuracy reached by the
electroweak (EW) precision measurements performed at the high-energy
$\ee$ colliders at LEP and SLC impose stringent constraints on the
coefficients of the operators entering the EW sector. The best studied
operators for constraining new physics (NP) are those arising from the
vector boson two-point functions \cite{oblique}, commonly referred to
as oblique or universal corrections. A little less prominent are the
specific left-handed (LH) contributions to the $\Zbb$ coupling
\cite{epsilonb}, which are known as vertex or non-universal
corrections. The tight experimental constraints \cite{ewpm} on the
three universal parameters $\eps_1$ ($T$), $\eps_2$ ($U$), and
$\eps_3$ ($S$), and the single non-universal parameter $\eps_b$
($\gam_b$) pose serious challenges for any conceivable extension of
the SM close to the EW scale.

Other severe constraints concern extra sources of flavor and $\CP$
violation that represent a generic problem in many NP scenarios. In
recent years great experimental progress has come primarily from the
BaBar and Belle experiments running on the $\eetoU$ resonance, leading
not only to an impressive accuracy in the determination of the
Cabibbo-Kobayashi-Maskawa (CKM) parameters \cite{ckm} from the
analysis of the unitarity triangle (UT) \cite{Charles:2004jd,
  Ciuchini:2000de}, but also excluding the possibility of new generic
flavor-violating couplings at the $\TeV$ scale. The most pessimistic
yet experimentally well supported solution to the flavor puzzle is to
assume that all flavor and $\CP$ violation is governed by the known
structure of the SM Yukawa interactions. This assumption defines
minimal-flavor-violation (MFV) \cite{Chivukula:1987py, MFV,
Buras:2000dm} independently of the specific structure of the NP
scenario \cite{D'Ambrosio:2002ex}. In the case of a SM-like Higgs
sector the resulting effective theory allows one to study correlations
between $K$- and $B$-decays \cite{D'Ambrosio:2002ex, Buras:2003jf,
  Bobeth:2005ck} since, by virtue of the large top quark Yukawa
coupling, all flavor-changing effective operators involving external
down-type quarks are proportional to the same non-diagonal structure
\cite{D'Ambrosio:2002ex}.  The absence of new $\CP$ phases in the
quark sector does not bode well for a dynamical explanation of the
observed baryon asymmetry of the universe. By extending the notion of
MFV to the lepton sector \cite{Cirigliano:2005ck}, however,
baryogenesis via leptogenesis has been recently shown to provide a
viable mechanism \cite{MLFV}.

The purpose of this article is to point out that in MFV scenarios
there exists a striking correlation between the $\Ztobb$ pseudo
observables (POs) $\Rb$, $\Ab$, and $\AFB$ measured at high-energy
$\ee$ colliders and all $Z$-penguin dominated low-energy
flavor-changing-neutral-current (FCNC) processes, such as $\Kpnn$,
$\KLnn$, $\KLmm$, $\BXdsnn$, and $\Bdsmm$ just to name a
few.\footnote{Of course, $\epseps$, $\KLpill$, $\BXdsll$ and all
  exclusive $\btodsll$ transitions could be mentioned here too.} The
crucial observation in this respect is that in MFV there is in general
a intimate relation between the non-universal contributions to the
anomalous $\Zbb$ couplings and the corrections to the flavor
off-diagonal $\Zdjdi$ operators since, by construction, NP couples
dominantly to the third generation. In particular, all specific MFV
models discussed in the following share the latter feature: the
two-Higgs-doublet model (THDM) type I and II, the
minimal-supersymmetric SM (MSSM) with MFV \cite{MFV, Buras:2000dm},
all for small $\tan \beta$, the minimal universal extra dimension
(mUED) model \cite{Appelquist:2000nn}, and the littlest Higgs model
\cite{Arkani-Hamed:2002qy} with $T$-parity (LHT) \cite{tparity} and
degenerate mirror fermions \cite{Low:2004xc}. Note that we keep our
focus on the LH contribution to the $Z$-penguin amplitudes, and thus
restrict ourselves to the class of constrained MFV (CMFV)
\cite{Buras:2003jf, Blanke:2006ig} models, i.e., scenarios that
involve no new effective operators besides those already present in
the SM. As our general argument does not depend on the chirality of
the new interactions it also applies to right-handed (RH) operators,
though with the minor difficulty of the appearance of an additional
universal parameter. Such an extension which covers large $\tan \beta$
contributions arising in a more general framework of MFV
\cite{D'Ambrosio:2002ex} is left for further study.

This article is organized as follows. In the next section we give a
model-independent argument based on the small momentum expansion of
Feynman integrals that suggests that the differences between the
values of the non-universal $\Zbb$ vertex form factors evaluated
on-shell and at zero external momenta are small in NP models with
extra heavy degrees of freedom. The results of the explicit
calculations of the one-loop corrections to the non-universal LH
contributions to the anomalous $\Zbb$ coupling in the CMFV models we
examine confirm these considerations. They are presented in
\Sec{sec:calculations}. \Sec{sec:numerics} contains a numerical
analysis of the allowed range for the non-standard contribution to the
$Z$-penguin function $C$ following from the presently available
data. In this section also lower and upper bounds for the branching
ratios of several rare $K$- and $B$-decays within CMFV based on these
ranges are derived. Concluding remarks are given in
\Sec{sec:conclusions}. \Appsand{app:cfunctions}{app:input} collects
the analytic expressions for the non-universal contributions to the
renormalized LH $\Zbb$ vertex functions in the considered CMFV models
and the numerical input parameters.

\section{General considerations}
\label{sec:general}

The possibility that new interactions unique to the third generation
lead to a relation between the LH non-universal $\Zbb$ coupling and
the LH flavor non-diagonal $\Zdjdi$ operators has been considered in a
different context before \cite{Chanowitz:1999jj}. Whereas the former
structure is probed by the ratio of the width of the $Z$-boson decay
into bottom quarks and the total hadronic width, $\Rb$, the bottom
quark left-right asymmetry parameter, $\Ab$, and the forward-backward
asymmetry for bottom quarks, $\AFB$, the latter ones appear in FCNC
transitions involving $Z$-boson exchange.

In the effective field theory framework of MFV
\cite{D'Ambrosio:2002ex}, one can easily see how the LH non-universal
$\Zbb$ coupling and the LH flavor non-diagonal $\Zdjdi$ operators are
linked together. The only relevant dimension-six contributions
compatible with the flavor group of MFV stem from the $SU(2) \times
U(1)$ invariant operators
\beq \label{eq:zoperators}
\begin{aligned} 
{\cal O}_{\phi 1} & = i \left( {\bar Q}_L Y_U Y_U^\dagger \gamma_\mu Q_L
\right) \phi^\dagger D^\mu \phi \, , \\
{\cal O}_{\phi 2} & = i \left( {\bar Q}_L Y_U Y_U^\dagger \tau^a
\gamma_\mu Q_L\right) \phi^\dagger \tau^a D^\mu \phi \, ,
\end{aligned}
\eeq 
that are built out of the LH quark doublets $Q_L$, the Higgs field
$\phi$, the up-type Yukawa matrices $Y_U$, and the $SU(2)$ generators
$\tau^a$. After EW symmetry breaking these operators are responsible
both for the non-universal $\Zbb$ coupling ($i = j = b$) and the
effective $\Zdjdi$ vertex ($i \not = j$). Since all SM up-type quark
Yukawa couplings $y_{u_i}$ except the one of the top, $y_t$, are
small, one has $(Y_U Y_U^\dagger)_{ji} \approx y_t^2 V_{tj}^\ast
V_{ti}$ so that only the top quark contribution to \Eq{eq:zoperators}
matters in practice.

That there exists a close relation is well-known in the case of the SM
where the same Feynman diagrams responsible for the enhanced top
correction to the anomalous $\Zbb$ coupling also generate the $\Zdjdi$
operators. In fact, in the limit of infinite top quark mass the
corresponding amplitudes are identical up to trivial CKM factors. Yet
there is a important difference between them. While for the physical
$\Ztobb$ decay the diagrams are evaluated on-shell, in the case of the
low-energy $\Ztodjdi$ transitions the amplitudes are Taylor-expanded
up to zeroth order in the off-shell external momenta before performing
the loop integration. As far as the momentum of the $Z$-boson is
concerned the two cases correspond to the distinct points $q^2 =
\MZ^2$ and $q^2 = 0$ in phase-space.

\begin{figure}[t!]
\begin{center}
\makebox{\hspace{+0mm} \scalebox{0.725}{\includegraphics{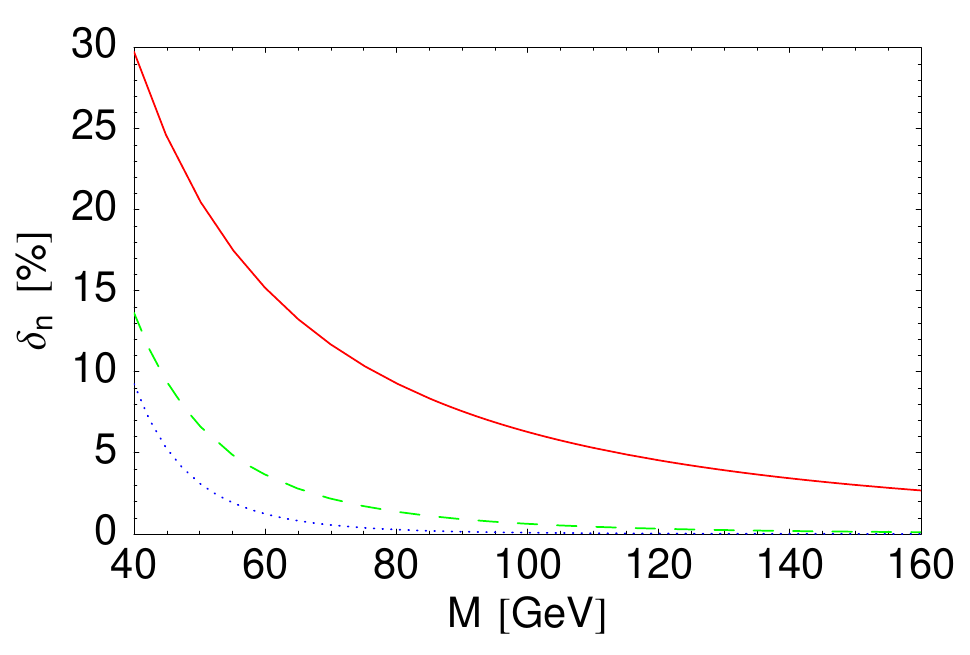}}}
\makebox{\hspace{-2mm} \scalebox{0.735}{\includegraphics{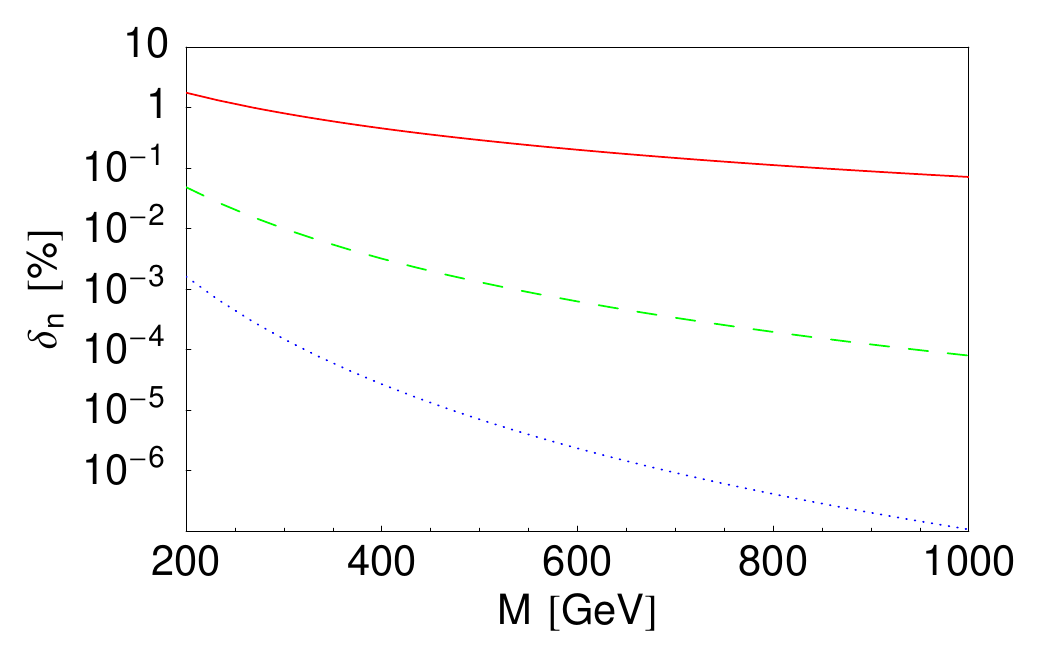}}}
\end{center}
\vspace{-6mm}
\caption{\sf Relative deviations $\delta_n$ for low (upper panel) and
  high values (lower panel) of $M$. The solid, dashed, and dotted
  curve correspond to $n = 1,2,$ and $3$, respectively. In obtaining
  the numerical values we have set $\MZ = 91 \, \GeV$ and $\mt = 165
  \, \GeV$. See text for details.}
\label{fig:sme}
\end{figure}

Observe that there is a notable difference between the small momentum
expansion and the heavy top quark mass limit. In the former case one
assumes $q^2 \ll \MW^2, \mt^2$ while in the latter case one has $q^2,
\MW^2 \ll \mt^2$. This difference naturally affects the convergence
behavior of the series expansions. While the heavy top quark mass
expansion converges slowly in the case of the non-universal one-loop
SM corrections to the $\Zbb$ vertex \cite{zbb}, we will demonstrate
that the small momentum expansion is well behaved as long as the
masses of the particles propagating in the loop are not too small,
i.e., in or above the hundred $\GeV$ range.     

The general features of the small momentum expansion of the one-loop
$\Zbb$ vertex can be nicely illustrated with the following simple but 
educated example. Consider the scalar integral 
\beq \label{eq:c0}
C_0 = \f{m_3^2}{i \pi^2} \int \! \f{d^4 l}{D_1 D_2 D_3} \, ,
\hspace{5mm} D_i \equiv (l + p_i)^2 - m_i^2 \, ,
\eeq
with $p_3 = 0$. Note that we have set the space-time dimension to four
since the integral is finite and assumed without loss of generality
$m_3 \neq 0$.

In the limit of vanishing bottom quark mass one has for the
corresponding momenta $p^2_1 = p^2_2 = 0$. The small momentum expansion
of the scalar integral $C_0$ then takes the form  
\beq \label{eq:sme}
C_0 = \sum_{n = 0}^\infty a_n \left ( \f{q^2}{m_3^2} \right )^n \, ,
\eeq
with $q^2 = (p_1 - p_2)^2 = -2 \hspace{0.2mm}  p_1 \! \cdot \! p_2$. The
expansion coefficients $a_n$ are given by \cite{Fleischer:1994ef}
\beq \label{eq:an}
a_n = \f{(-1)^n}{(n + 1)!} \sum_{l = 0}^n \begin{pmatrix} n \\
  l \end{pmatrix} \f{x_1^l}{l!}  \f{\partial^l}{\partial x_1^l}
\f{\partial^n}{\partial x_2^n} g(x_1, x_2) \, ,
\eeq
where
\beq \label{eq:gx1x2}
g(x_1, x_2) = \f{1}{x_1 - x_2} \left ( \f{x_1 \ln x_1}{1 - x_1} -
  \f{x_2 \ln x_2}{1 - x_2} \right ) \, ,
\eeq
and $x_i \equiv m_i^2/m_3^2$. Notice that in order to properly
generate the expansion coefficients $a_n$ one has to keep $x_1$ and
$x_2$ different even in the zero or equal mass case. The corresponding
limits can only be taken at the end.     

In order to illustrate the convergence behavior of the small momentum
expansion of the scalar integral in \Eq{eq:sme} for on-shell
kinematics, we confine ourselves to the simplified case $m_1 = m_2 =
M$ and $m_3 = \mt$. We define      
\beq \label{eq:deltan}
\delta_n \equiv a_n \left ( \f{\MZ^2}{\mt^2} \right )^n \left (
  \sum_{l = 0}^{n - 1} a_l \left ( \f{\MZ^2}{\mt^2} \right )^l \right 
)^{-1} \, , 
\eeq
for $n = 1, 2, \ldots \,$. The $M$-dependence of the relative
deviations $\delta_n$ is displayed in \Fig{fig:sme}. We see that while
for values of $M$ much below $\mt$ higher order terms in the small
momentum expansion have to be included in order to approximate the
exact on-shell result accurately, in the case of $M$ larger than $\mt$
already the first correction is small and higher order terms are
negligible. For the two reference scales $M = 80 \, \GeV$ and $M = 250
\, \GeV$ one finds for the first three relative deviations $\delta_n$
numerically $+9.3 \%$, $+1.4 \%$, and $+0.3 \%$, and $+1.1 \%$, $+0.02
\%$, $+0.00004 \%$, respectively.   

It should be clear that the two reference points $M = 80 \, \GeV$ and
$M = 250 \, \GeV$ have been picked for a reason. While the former
describes the situation in the SM, i.e., the exchange of two pseudo
Goldstone bosons and a top quark in the loop, the latter presents a
possible NP contribution arising from diagrams containing two heavy
scalar fields and a top quark. The above example indicates that the
differences between the values of the non-universal $\Zbb$ vertex form
factors evaluated on-shell and at zero external momenta are in general
much less pronounced in models with extra heavy degrees of freedom
than in the SM. In view of the fact that this difference amounts to a
modest effect of around $-30 \%$ in the SM \cite{zbb}, it is
suggestive to assume that the scaling of NP contributions to the
non-universal parts of the $\Zbb$ vertex is in general below the $\pm
10 \%$ level. This model-independent conclusion is well supported by
the explicit calculations of the one-loop corrections to the specific
LH contribution to the anomalous $\Zbb$ coupling in the CMFV versions
of the THDM, the MSSM, the mUED, and the LHT model presented in the
next section.

We would like to stress that our general argument does not depend on
the chirality of possible new interactions as it is solely based on
the good convergence properties of the small momentum expansion of the
relevant vertex form factors. Thus we expect it to hold in the case of
RH operators as well. Notice that the assumption of MFV does not play
any role in the flow of the argument itself as it is exerted only at
the very end in order to establish a connection between the $\Zbb$ and
$\Zdjdi$ vertices evaluated at zero external momenta by a proper
replacement of CKM factors. Therefore it does not seem digressive to
anticipate similar correlations between the flavor diagonal and
off-diagonal $Z$-penguin amplitudes in many beyond-MFV scenarios in
which the modification of the flavor structure is known to be
dominantly non-universal, i.e., connected to the third generation. See
\cite{thirdgeneration} for a selection of theoretically well-motivated
realizations. These issues warrant a detailed study.

\section{Model calculations}
\label{sec:calculations}

The above considerations can be corroborated in another, yet
model-dependent way by calculating explicitly the difference between
the value of the LH $\Zdjdi$ vertex form factor evaluated on-shell and
at zero external momenta. In the following this will be done in four
of the most popular, consistent, and phenomenologically viable
scenarios of CMFV, i.e., the THDM, the MSSM, both for small $\tan
\beta$, the mUED, and the LHT model, the latter in the case of
degenerate mirror fermions. All computations have been performed in
the on-shell scheme employing the 't Hooft-Feynman gauge. The actual
calculations were done with the help of the packages {\it FeynArts}
\cite{Hahn:2000kx} and {\it FeynCalc} \cite{Mertig:1990an}, and {\it
  LoopTools} \cite{Hahn:1998yk} and {\it FF}
\cite{vanOldenborgh:1990yc} for numerical evaluation.

\begin{figure}[!t]
\begin{center}
\hspace{-7.5mm}
\scalebox{0.8}{\includegraphics{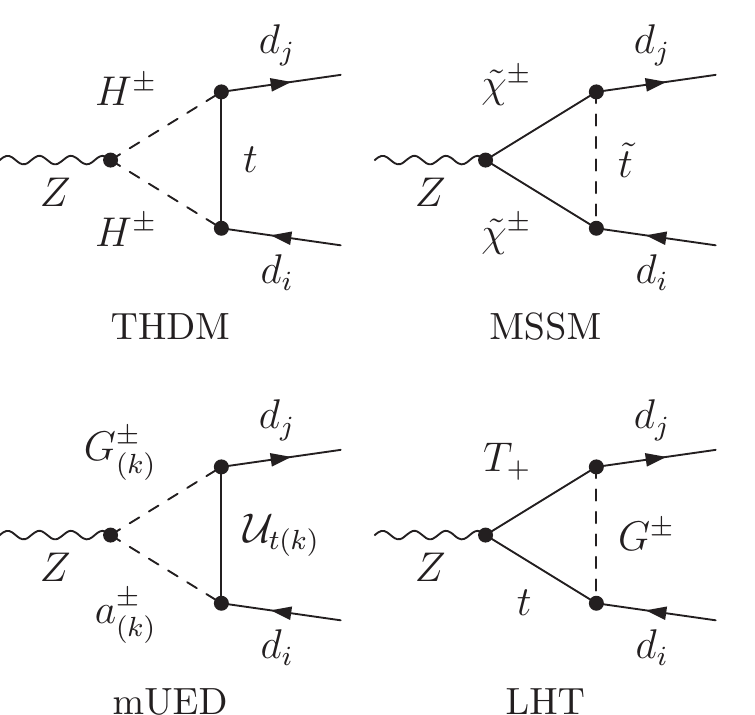}}
\end{center}
\vspace{-5mm}
\caption{\sf Examples of one-loop vertex diagrams that result in a  
  non-universal correction to the $\Ztodjdi$ transition in assorted 
  NP scenarios with CMFV. See text for details. }  
\label{fig:cmfv}
\end{figure}

Before presenting our results\footnote{The analytic expressions for
  the renormalized $\Zdjdi$ vertex functions in the considered CMFV
  models are collected in \App{app:cfunctions}.} we collect a couple
of definitions to set up our notation. In the limit of vanishing
bottom quark mass, possible non-universal NP contributions to the
renormalized LH off-shell $\Zdjdi$ vertex can be written as
\beq \label{eq:zdjdi}
\Gamma_{ji}^{\rm NP} = \f{\GF}{\sqrt{2}} \f{e}{\pi^2} \MZ^2
\f{\cw}{\sw} V_{tj}^\ast V_{ti} C_{\rm NP} (q^2) \bar{d_j}_\sL
\gamma_\mu {d_i}_\sL Z^\mu \, ,   
\eeq
where $i = j = b$ and $i \neq j$ in the flavor diagonal and
off-diagonal cases. $\GF$, $e$, $\sw$, and $\cw$ denote the Fermi
constant, the electromagnetic coupling constant, the sine and cosine
of the weak mixing angle, respectively, while $V_{ij}$ are the
corresponding CKM matrix elements and the subscript $L$ indicates that
the interactions involve LH down-type quark fields only.

As a measure of the relative difference between the complex valued
form factor $C_{\rm NP} (q^2)$ evaluated on-shell and at zero momentum 
we introduce  
\beq \label{eq:dcnp}
\delta C_{\rm NP} \equiv 1 - \f{\re \, C_{\rm NP} (q^2 = 0)}{\re \,
  C_{\rm NP} (q^2 = \MZ^2)} \, .  
\eeq  

In the THDM with vanishing tree-level FCNCs, the only additional
contribution to the $\Ztodjdi$ transitions with respect to the SM
comes from loops containing charged Higgs bosons, $H^\pm$, and top
quarks, $t$. An example of such a contribution is shown on the top
left-hand side of \Fig{fig:cmfv}. The correction depends on the mass
of the charged Higgs boson, $\MHpm$, and on the ratio of the vacuum
expectation value of the Higgs doublets, $\tan \beta$. Models of type
I and II differ in the way quarks couple to the Higgs doublets: in the
type I scenario both the masses of down- and up-type quarks are
generated by one of the doublets, like in the SM, while in the type II
theory one of the doublets generates the down-type and the second one
generates the up-type masses, like in the MSSM. In our case only the
coupling to the top quark is relevant, so that we do not need to
actually distinguish between types I and II.

\begin{figure}[!t]
\begin{center}
\makebox{\hspace{+0mm} \scalebox{0.725}{\includegraphics{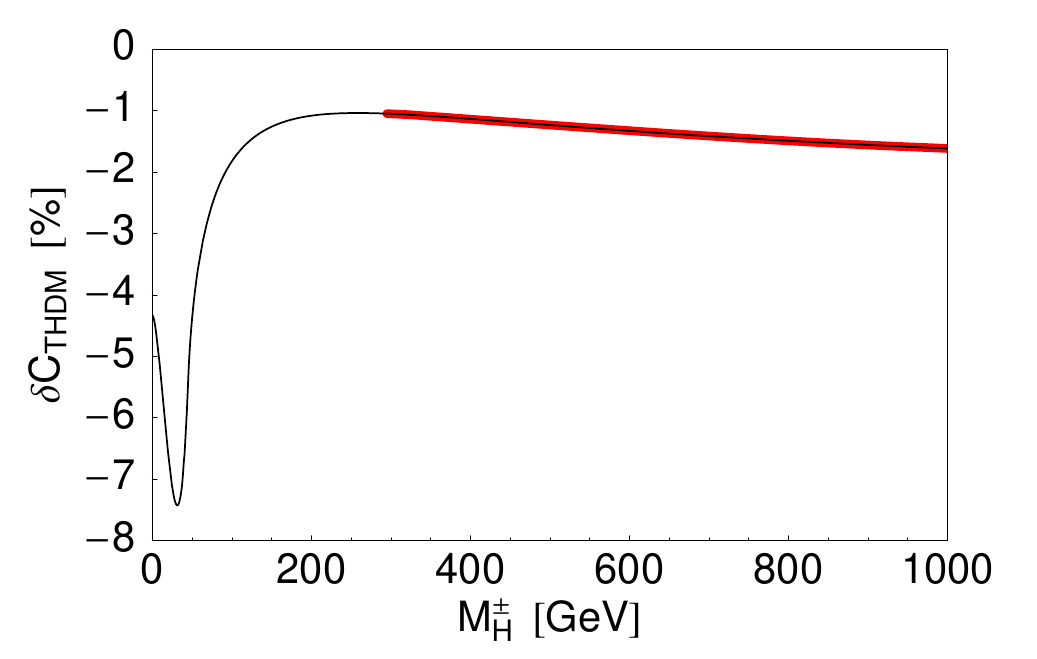}}}
\makebox{\hspace{-3mm} \scalebox{0.740}{\includegraphics{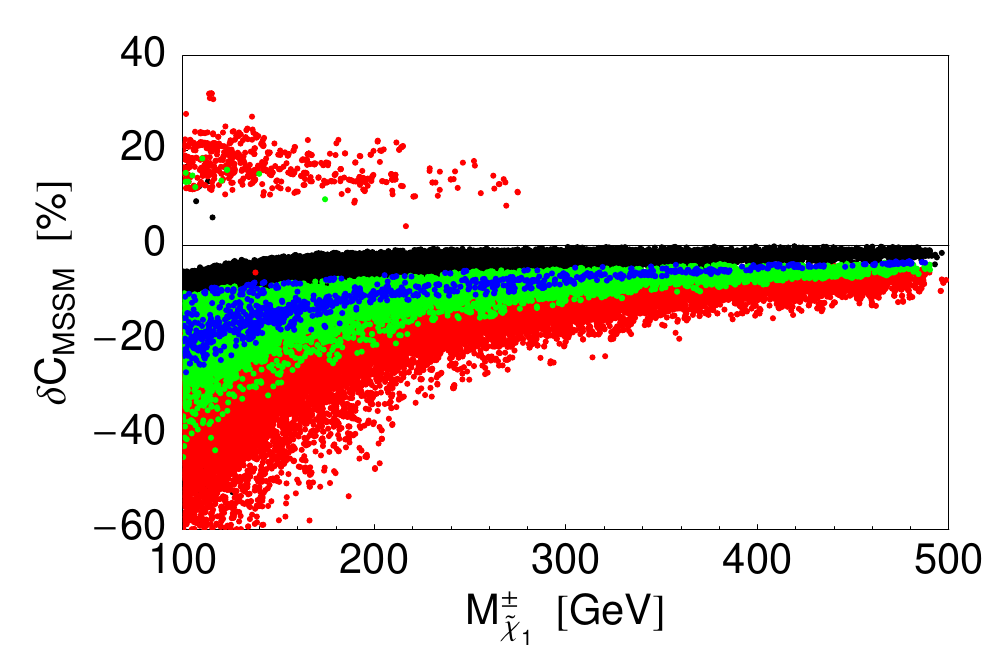}}}
\makebox{\hspace{-2mm} \scalebox{0.725}{\includegraphics{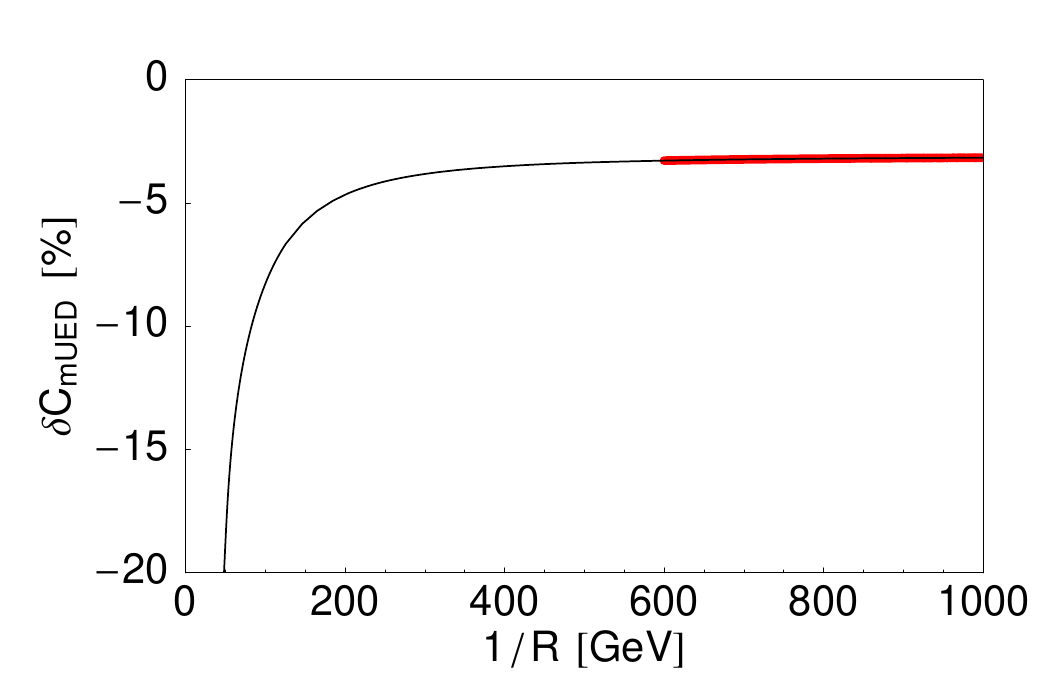}}}
\makebox{\hspace{-6mm} \scalebox{0.725}{\includegraphics{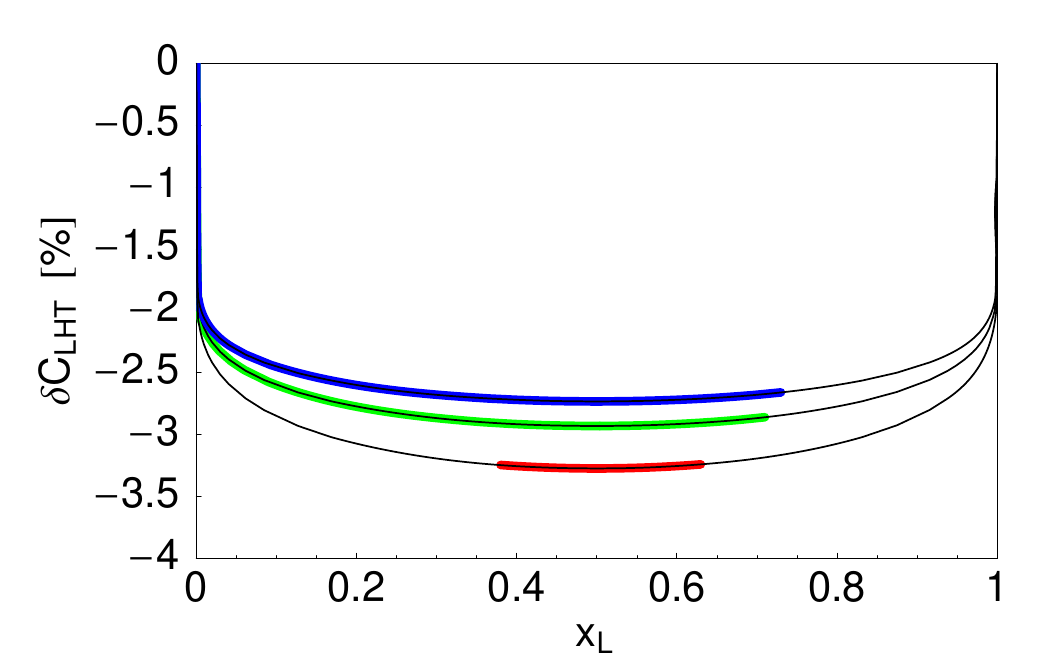}}}
\end{center}
\vspace{-6mm}
\caption{\sf Relative difference $\delta C_{\rm NP}$ in the THDM, the
  MSSM, the mUED, and the LHT model as a function of $\MHpm$, $\mcha$,
  $1/R$, and $x_L$. The allowed parameter regions after applying
  experimental and theoretical constraints are indicated by the
  colored (grayish) bands and points. See text for details.}
\label{fig:scalings}
\end{figure}

To find $\delta C_{\rm THDM}$ we have computed analytically the
one-loop charged Higgs corrections to \Eq{eq:zdjdi} reproducing the
result of \cite{Denner:1991ie}. The analytic expression for $C_{\rm
  THDM} (q^2)$ can be found in \Eq{eq:CTHDM}. The dependence of
$\delta C_{\rm TDHM}$ on $\MHpm$ can be seen in the first panel of
\Fig{fig:scalings}. The red (gray) band underlying the solid black
curve shows the part of the parameter space satisfying the lower bound
$\MHpm \gtrsim 295 \, \GeV$ following from $\BXsga$ in the THDM of
type II using the most recent SM prediction \cite{bsg}. This $\tan
\beta$ independent bound is much stronger than the one from the direct
searches at LEP corresponding to $\MHpm > 78.6 \, \GeV$
\cite{Yao:2006px}, and than the indirect lower limits from a number of
other processes. In model I, the most important constraint on $\MHpm$
comes from $\Rb$ \cite{Haber:1999zh}. As the corresponding bound
depends strongly on $\tan \beta$ we do not include it in the
plot. While the decoupling of $\delta C_{\rm THDM}$ occurs slowly, we
find that the maximal allowed relative suppression of $\re \, C_{\rm
  THDM} (q^2 = \MZ^2)$ with respect to $\re \, C_{\rm THDM} (q^2 = 0)$
is below $2 \%$ and independent of $\tan \beta$, as the latter
dependence exactly cancels out in \Eq{eq:dcnp}. In obtaining the
numerical values for $\delta C_{\rm TDHM}$ we have employed $\MW = 80
\, \GeV$, $\MZ = 91 \, \GeV$, $\mt = 165 \, \GeV$, and $\sws =
0.23$. If not stated otherwise, the same numerical values will be used
in the remainder of this article. We assess the smallness of $\delta
C_{\rm THDM}$ as a first clear evidence for the correctness of our
general considerations.

In the case of the MSSM with conserved $R$-parity, we focus on the
most general realization of MFV compatible with renormalization group
(RG) invariance \cite{D'Ambrosio:2002ex}. In this scenario CKM-type
flavor- and $\CP$-violating terms appear necessarily in the down- and
up-type squark mass-squared matrices due to the symmetry principle
underlying the MFV hypothesis. The explicit form of the physical
up-type squark mass matrix used in our analysis is given in
\Eq{eq:Msup}. We assume universality of soft supersymmetry (SUSY)
breaking masses and proportionality of trilinear terms at the EW
scale,\footnote{If universality of soft SUSY breaking masses and
  proportionality of trilinear terms is assumed at some high-energy
  scale off-diagonal entries are generated by the RG running down to
  the EW scale. We ignore this possibility here.} so that neutralino
and gluino contributions to flavor-changing $\Ztodjdi$ transitions are
absent. This additional assumption about the structure of the soft
breaking terms in the squark sector has a negligible effect on the
considered FCNC processes \cite{Isidori:2006qy}.\footnote{In
  \cite{Altmannshofer:2007cs} it has been pointed out that in
  scenarios characterized by large values of the higgsino mass
  parameter, i.e., $|\mu| \approx 1 \, \TeV$, the MFV MSSM with small
  $\tan \beta$ is not necessarily CMFV due to the presence of
  non-negligible gluino corrections in $\Delta B = 2$ amplitudes. This
  observation is irrelevant for our further discussion.}  Moreover, in
the small $\tan \beta$ regime both neutralino and neutral Higgs
corrections to $\Ztobb$ turn out to be insignificant
\cite{Boulware:1991vp}. Therefore only SUSY diagrams involving
chargino, $\tilde{\chi}^\pm$, and stop, $\tilde{t}$, exchange are
relevant here. An example of such a contribution can be seen on the
top right side of \Fig{fig:cmfv}. A noticeable feature in the chosen
setting is that large left-right mixing can occur in the stop sector,
leading to both a relatively heavy Higgs in the range $120 \, \GeV
\lesssim \mh \lesssim 135 \, \GeV$ and a stop mass eigenstate, say,
$\tilde{t}_1$, possibly much lighter than the remaining squarks.  Such
a scenario corresponds to the ``golden region'' of the MSSM, where all
experimental constraints are satisfied and fine-tuning is minimized
\cite{Perelstein:2007nx}. For what concerns the other sfermions we
neglect left-right mixing and assume that all squarks and sleptons
have a common mass $\msq$ and $\msl$, respectively.\footnote{A strict
  equality of left-handed squark masses is not allowed due to the
  different D-terms in the down- and up-type squark sector. For our
  purposes this difference is immaterial.}

In order to find the complete MSSM correction $\delta C_{\rm MSSM}$ we
have calculated analytically the one-loop chargino-up-squark
corrections to \Eq{eq:zdjdi} and combined it with the charged Higgs
contribution. Our result for $C_{\tilde{\chi}^\pm} (q^2)$ agrees with
the one of \cite{Boulware:1991vp}\footnote{The last equation in this
  article has a typographic error. The Passarino-Veltman function
  $C_{11}$ should read $C_{12}$.} and is given in \Eq{eq:CCha}. The
region of parameters in which the SUSY corrections to the LH $\Zdjdi$
vertices are maximal corresponds to the case of a light stop and
chargino. In our numerical analysis we therefore focus on these
scenarios. We allow the relevant MSSM parameters to float freely in
the ranges $2 < \tan \beta < 6$, $|\mu| < 500 \, \GeV$, and $M < 1 \,
\TeV$ for $M = M_H^\pm, M_2, \mstone, \msq, m_{\tilde{l}}$. The value
of the trilinear coupling $A_u$ is computed from each randomly chosen
set of parameters $\mu$, $\mstone$, and $\msq$. The calculation of
$\btosgamma$ and $\btosll$ that is used to constrain the parameter
space introduces also a dependence on the gluino mass $\mgl$. We
choose to vary $\mgl$ in the range $241 \, \GeV < \mgl < 1 \, \TeV$
\cite{Abazov:2006bj}.

The MSSM parameter space is subject to severe experimental and
theoretical constraints. We take into account the following lower
bounds on the particle masses \cite{Yao:2006px}: $\MHpm > 78.6 \,
\GeV$, $\mcha > 94 \, \GeV$, $\mstone > 95.7 \, \GeV$, $\msq > 99.5 \,
\GeV$, and $\msl > 73 \, \GeV$. In the considered parameter space the
requirement of the absence of color and/or charge breaking minima sets
a strong upper limit of around $3 \, \TeV$ on the absolute value of
$A_u$ \cite{ccbreak}. As far as the lightest neutral Higgs boson is
concerned, we ensure that $\mh > 114.4 \, \GeV$ \cite{ewpm}, including
the dominant radiative corrections \cite{mhsusy} to its tree-level
mass. Further restrictions that we impose on the SUSY parameter space
are the $\rho$ parameter \cite{rhosusy}, $\Rb$ \cite{Boulware:1991vp},
and the inclusive $\btosgamma$ \cite{bsgsusy} and $\btosll$
\cite{Bobeth:2004jz} branching fractions. To find the boundaries of
the allowed parameter space we perform an adaptive scan of the eight
SUSY variables employing the method advocated in \cite{adaptive}.

The dependence of $\delta C_{\rm MSSM}$ on the lighter chargino mass
$\mcha$ is illustrated in the second plot of
\Fig{fig:scalings}. Regions in the $\mcha$--$\, \delta C_{\rm MSSM}$
plane where the absolute value of the correction $\re \, C_{\rm MSSM}
(q^2 = 0)$ amounts to at least $2 \%$, $4 \%$, $6 \%$, and $10 \%$ of
the SM value $\re \, C_{\rm SM} (q^2 = 0)$ are indicated by the red
(gray), green (light gray), blue (dark gray), and black points,
respectively. No constraints are imposed for the black points while
the colored (grayish) ones pass all the collider and low-energy
constraints mentioned above. 

Three features of our numerical explorations deserve special
mention. First, the maximal allowed relative size of the correction
$\re \, C_{\rm MSSM} (q^2 = 0)$ amounts to less than $^{+ 9 \%}_{-6
  \%}$ of $\re \, C_{\rm SM} (q^2 = 0)$. Second, the magnitude of the
possible deviation $\delta C_{\rm MSSM}$ is strongly anti-correlated
with the absolute size of $\re \, C_{\rm MSSM} (q^2 = 0)$. While small
corrections $\re \, C_{\rm MSSM} (q^2 = 0)$ allow for large values of
$\delta C_{\rm MSSM}$ the latter difference decreases rapidly with
increasing $\re \, C_{\rm MSSM} (q^2 = 0)$. Third, the correction
$\delta C_{\rm MSSM}$ decouples quickly for heavy charginos. These
features imply that $\delta C_{\rm MSSM}$ is small if one requires
$(i)$ the relative size of $\re \, C_{\rm MSSM} (q^2 = 0)$ to be
observable, i.e., to be bigger than the SM uncertainty of the
universal $Z$-penguin function\footnote{The overall uncertainty of
  $\re \, C_{\rm SM} (q^2 = 0)$ amounts to around $\pm 3 \%$. It is in
  equal shares due to the parametric error on the top quark mass, the
  matching scale uncertainty in the next-to-leading order result
  \cite{Buchalla:1992zm}, and two-loop EW effects that are only partly
  known \cite{Buchalla:1997kz}.} and $(ii)$ the chargino mass $\mcha$
to be not too light. For example, all allowed points satisfy $|\delta
C_{\rm MSSM}| < 0.01$ if one demands $|\re \, C_{\rm MSSM} (q^2 =
0)/\re \, C_{\rm SM} (q^2 = 0)| > 0.05$ and $\mcha > 300 \ \GeV$. On
the other hand, if the masses of the lighter chargino and stop both
lie in the hundred $\GeV$ range, $\delta C_{\rm MSSM}$ frequently
turns out to be larger than one would expected on the basis of our
model-independent considerations.

The large corrections $\delta C_{\rm MSSM}$ can be traced back to the
peculiar structure of the form factor $C_{\tilde{\chi}^\pm}
(q^2)$. While in the limit of vanishing external $Z$-boson momentum
the first three terms in \Eq{eq:CCha} all approach a constant value
the fourth one scales like $q^2/M_{\rm SUSY}^2$ with $M_{\rm SUSY} =
{\rm min}(\mcha, \mstone)$. Naively, one thus would expect the general
argument given in the last section to hold. Yet for large left-right
mixing in the stop sector, which permits a relatively heavy Higgs mass
of $\mh \gtrsim 120 \, \GeV$, it turns out that the first three
contributions tend to cancel each other and, in turn, the size of
$\delta C_{\rm MSSM}$ is controlled by the fourth term. Then $\delta
C_{\rm MSSM} \propto \MZ^2/M_{\rm SUSY}^2$ and the correction $\delta
C_{\rm MSSM}$ can be sizable if $M_{\rm SUSY}$ is close to the EW
scale. The observed numerical cancellation also explains why $\delta
C_{\rm MSSM}$ is typically large if $\re \, C_{\rm MSSM} (q^2 = 0)$ is
small and vice versa. It should be clear, however, that the large
deviation $\delta C_{\rm MSSM}$ are ultimately no cause of concern,
because $|\re \, C_{\rm MSSM} (q^2 = 0)/\re \, C_{\rm SM} (q^2 = 0)|$
itself is always below $10 \%$. In consequence, the model-independent
bound on the NP contribution to the universal $Z$-penguin function
that we will derive in the next section does hold in the case of the 
CMFV MSSM.

Among the most popular non-SUSY models in question is the model of
Appelquist, Cheng, and Dobrescu (ACD) \cite{Appelquist:2000nn}. In the
ACD framework the SM is extended from four-dimensional Minkowski
space-time to five dimensions and the extra space dimension is
compactified on the orbifold $S^1/Z_2$ in order to obtain chiral
fermions in four dimensions. The five-dimensional fields can
equivalently be described in a four-dimensional Lagrangian with heavy
Kaluza-Klein (KK) states for every field that lives in the fifth
dimension or bulk. In the ACD model all SM fields are promoted to the
bulk and in order to avoid large FCNCs tree-level boundary fields and
interactions are assumed to vanish at the cut-off
scale.\footnote{Boundary terms arise radiatively \cite{uedbrane}. They
  effect the $\Ztodjdi$ amplitude first at the two-loop level. Since
  we perform a leading order analysis in the ACD model its consistent
  to neglect these effects.} A remnant of the translational symmetry
after compactification leads to KK-parity. This property implies, that
KK states can only be pair-produced, that their virtual effect comes
only from loops, and causes the lightest KK particle to be stable,
therefore providing a viable dark matter candidate \cite{ueddm}.

In the following we will assume vanishing boundary terms at the
cut-off scale and that the ultraviolet (UV) completion does not
introduce additional sources of flavor and $\CP$ violation beyond the
ones already present in the model. These additional assumptions define
the mUED model which then belongs to the class of CMFV scenarios. The
one-loop correction to $\Gamma_{ji}^{\rm mUED}$ is found from diagrams
containing apart from the ordinary SM fields, infinite towers of the
KK modes corresponding to the $W$-boson, $W^\pm_{(k)}$, the pseudo
Goldstone boson, $G^\pm_{(k)}$, the $SU (2)$ quark doublets, ${\cal
  Q}_{q (k)}$, and the $SU (2)$ quark singlets, ${\cal U}_{q
  (k)}$. Additionally, there appears a charged scalar, $a^\pm_{(k)}$,
which has no counterpart in the SM. A possible diagram involving such
a KK excitation is shown on the lower left side in
\Fig{fig:cmfv}. Since at leading order the $\Ztodjdi$ amplitude turns
out to be cut-off independent the only additional parameter entering
$\Gamma_{ji}^{\rm mUED}$ relative to the SM is the inverse of the
compactification radius $1/R$. The analytic expression for $C_{\rm
  mUED} (q^2)$ can be found in \Eq{eq:CACD}.

For a light Higgs mass of $\mh = 115 \, \GeV$ a careful analysis of
oblique corrections \cite{Gogoladze:2006br} gives a lower bound of
$1/R \gtrsim 600 \, \GeV$, well above current collider limits
\cite{acdcollider}. With increasing Higgs mass this constraint relaxes
significantly leading to $1/R \gtrsim 300 \, \GeV$
\cite{Gogoladze:2006br, Appelquist:2002wb}. Other constraints on $1/R$
that derive from $\Rb$ \cite{Oliver:2002up}, the muon anomalous
magnetic moment \cite{Appelquist:2001jz}, and flavor observables
\cite{Buras:2002ej, Buras:2003mk, acdflavor} are in general weaker. An
exception is the inclusive $\BXsga$ branching ratio. Since the SM
prediction \cite{bsg} is now lower than the experimental world average
by more than $1 \mysigma$ and the one-loop KK contributions interfere
destructively with the SM $\btosgamma$ amplitude \cite{Buras:2003mk,
  Agashe:2001xt}, $\BXsga$ provides at leading order the lower bound
$1/R \gtrsim 600 \, \GeV$ independent from the Higgs mass
\cite{Haisch:2007vb}. The $1/R$ dependence of $\delta C_{\rm mUED}$ is
displayed in the third plot of \Fig{fig:scalings}. In the range of
allowed compactification scales, indicated by the red (gray) stripe,
the suppression of $\re \, C_{\rm mUED}(q^2 = \MZ^2)$ compared to $\re
\, C_{\rm mUED} (q^2 = 0)$ amounts to less than $5 \%$, the exact
value being almost independent of $1/R$. This lends further support to
the conclusion drawn in the last section. We finally note that our new
result for $C_{\rm mUED} (q^2)$ coincides for $q^2 = 0$ with the
one-loop KK contribution to the $Z$-penguin function calculated in
\cite{Buras:2002ej}.

Another phenomenologically very promising NP scenario is the LHT
model. Here the Higgs is a pseudo Goldstone boson arising from the
spontaneous breaking of an approximate global $SU(5)$ symmetry down to
$SO(5)$ \cite{Arkani-Hamed:2002qy} at a scale $f$. To make the
existence of new particle in the $1 \, \TeV$ range consistent with
precision EW data, an additional discrete $Z_2$ symmetry called
$T$-parity \cite{tparity}, is introduced, which as one characteristic
forbids tree-level couplings that violate custodial $SU(2)$ symmetry.
In the fermionic sector, bounds on four fermion operators demand for a
consistent implementation of this reflection symmetry the existence of
a copy of all SM fermions, aptly dubbed mirror fermions
\cite{Low:2004xc}. The theoretical concept of $T$-parity and its
experimental implications resemble the one of $R$-parity in SUSY and
KK-parity in universal extra dimensional theories.

Unless their masses are exactly degenerate, the presence of mirror
quarks leads in general to new flavor- and $\CP$-violating
interactions. In order to maintain CMFV we are thus forced to assume
such a degeneracy here. In this case contributions from particles that
are odd under $T$-parity vanish due to the Glashow-Iliopoulos-Maiani
(GIM) mechanism \cite{Glashow:1970gm}, and the only new particle that
affects the $\Ztodjdi$ transition in a non-universal way is a $T$-even
heavy top, $T_+$. A sample diagram involving such a heavy top, its
also $T$-even partner, i.e., the top quark $t$, and a pseudo Goldstone
field, $G^\pm$, is shown on the lower right-hand side of
\Fig{fig:cmfv}. In turn, $\Gamma_{ji}^{\rm LHT}$ depends only on the
mass of the heavy quark $T_+$, which is controlled by the size of the
top Yukawa coupling, by $f$, and the dimensionless parameter $x_L
\equiv \lambda_1^2/(\lambda_1^2 + \lambda_2^2)$. Here $\lambda_1$ is
the Yukawa coupling between $t$ and $T_+$ and $\lambda_2$ parametrizes
the mass term of $T_+$. In the fourth panel of \Fig{fig:scalings}, we
show from bottom to top $\delta C_{\rm LHT}$ as a function of $x_L$
for $f = 1, 1.5,$ and $2 \, \TeV$. The colored (grayish) bands
underlying the solid black curves correspond to the allowed regions in
parameter space after applying the constraints following from
precision EW data \cite{Hubisz:2005tx}. As NP effects in the quark
flavor sector of the LHT model with CMFV are generically small
\cite{lhtflavor, Blanke:2006eb}, they essentially do not lead to any
restrictions. We find that the maximal allowed suppression of $\re \,
C_{\rm LHT} (q^2 = \MZ^2)$ with respect to $\re \, C_{\rm LHT} (q^2 =
0)$ is slightly bigger than $3 \%$. This feature again confirms our
general considerations. Our new result for $C_{\rm LHT} (q^2)$ given
in \Eq{eq:CLHT} resembles for $q^2 = 0$ the analytic expression of the
one-loop correction to the low-energy $Z$-penguin function calculated
in \cite{Blanke:2006eb}. Taking into account that the latter result
corresponds to unitary gauge while we work in 't Hooft-Feynman gauge
is essential for this comparison. In particular, in our case no UV
divergences remain after GIM, as expected on general grounds
\cite{Bardeen:2006wk}.

At this point a further comment concerning gauge invariance is in
order. It is well known that only a proper arrangement of, say,
$\eeff$, including all contributions related to the $Z$-boson, purely
EW boxes, and the photon, is gauge invariant at a given order in
perturbation theory. In flavor physics such a gauge independent
decomposition \cite{Buchalla:1990qz} is provided by the combinations
$\X \equiv \C + \Bnn$, $\Y \equiv \C + \Bll$, and $\Z \equiv \C +
\D/4$ of Inami-Lim functions \cite{Inami:1980fz}. Given the
normalization of \Eq{eq:zdjdi}, NP contributions to the universal
$Z$-penguin function $\C$ are characterized by $\re \, C_{\rm NP} (q^2
= 0)$ in our notation, while $\Bnn$ and $\Bll$ represent the
contribution of EW boxes with neutrino and charged lepton pairs in the
final state. $\D$ stems from the off-shell part of the magnetic photon
penguin amplitude. Since we want to relate in a model-independent way
observables derived from $\eeff$ to observables connected with the
$\didjnn$ and $\didjll$ transitions, we also have to worry about the
potential size of corrections that are not associated with the
$Z$-boson.

At the $Z$-pole, the total cross-section of $\eeff$ is completely
dominated by $Z$-boson exchange. While purely EW boxes are vanishingly 
small, the bulk of the radiative corrections necessary to interpret
the measurements are QED effects. It is important to realize that
these QED corrections are essentially independent of the EW ones, and
therefore allow the anomalous $\Zbb$ couplings to be extracted from 
the data in a model-independent manner. Certain SM assumptions are
nevertheless employed when extracting and interpreting the couplings,
but considerable effort \cite{ewpm} has been expended to make the
extraction of the POs $\Rb$, $\Ab$, and $\AFB$ as model-independent as
possible, so that the meanings of theory and experiment remain distinct.   
   
In the case of the $\didjnn$ and $\didjll$ observables theoretical
assumptions about the size of the EW boxes are unfortunately
indispensable. Our explicit analysis of the considered CMFV models
reveals the following picture. In the THDM, the NP contributions
$\Delta \Bnn \equiv \Bnn - \Bnn_{\rm SM}$ and $\Delta \Bll \equiv \Bll
- \Bll_{\rm SM}$ vanish identical \cite{Bobeth:2001jm}, while their
relative sizes compared to the corresponding SM contributions amount
to at most $^{+1}_{-11} \%$ and $^{+18}_{-5} \%$ in the MSSM
\cite{Buras:2000qz} and less than $+1 \%$ in both the mUED scenario
\cite{Buras:2002ej} and the LHT model \cite{Blanke:2006eb}. The
numbers for $\Delta \Bnn$, $\Delta \Bll$, and $\Delta C$ quoted here
and in the following refer to the 't Hooft-Feynman gauge. Moreover,
contributions to the EW boxes are found to be generically suppressed
by at least two inverse powers of the scale of NP using naive
dimensional analysis \cite{Buras:1999da}. In view of this, the
possibility of substantial CMFV contributions to the EW boxes seems
rather unlikely. The actual size of the NP contribution $\Delta \D
\equiv \D -\D_{\rm SM}$ to the off-shell magnetic photon penguin
function $\D$ has essentially no impact on our conclusions. The
treatment of $\Delta \Bnn$, $\Delta \Bll$, and $\Delta \D$ in our
numerical analysis will be discussed in the next section.

\section{Numerical analysis}
\label{sec:numerics}
   
Our numerical analysis consists of three steps. First we determine the
CKM parameters $\bar{\rho}$ and $\bar{\eta}$ from an analysis of the
universal UT \cite{Buras:2000dm}.\footnote{If the unitarity of the $3
  \times 3$ CKM matrix is relaxed sizable deviations from $V_{tb}
  \simeq 1$ are possible \cite{Alwall:2006bx}. We will not consider
  this possibility here since it is not covered by the MFV hypothesis
  which requires that all flavor and $\CP$ violation is determined by
  the structure of the ordinary $3 \times 3$ SM Yukawa couplings
  \cite{D'Ambrosio:2002ex}.} The actual analysis is performed with a
customized version of the CKMfitter package
\cite{Charles:2004jd}. Using the numerical values of the experimental
and theoretical parameters collected in \App{app:input} we find
\beq \label{eq:uutfit} \bar{\rho} = 0.160 \pm 0.031 \, , \hspace{5mm}
\bar{\eta} = 0.326 \pm 0.012 \, .  \eeq The given central values are
highly independent of $\mt$, but depend mildly on the hadronic
parameter $\xi \equiv (f_{B_s} \hat{B}_{B_s}^{1/2})/(f_{B_d}
\hat{B}_{B_d}^{1/2})$ determined in lattice QCD. Since in our approach
theoretical parameter ranges are scanned, the quoted $68 \%$
confidence levels (${\rm CL}$s) should be understood as lower bounds,
i.e., the range in which the quantity in question lies with a
probability of at least $68 \%$. The same applies to all ${\rm CL}$s
and probability regions given subsequently.

In the second step, we determine the allowed ranges of $\Delta C$ and
the NP contribution $\Delta C_7^{\rm eff} \equiv C_7^{\rm eff} (\mb) -
C_{7 \, {\rm SM}}^{\rm eff} (\mb)$ to the effective on-shell magnetic
photon penguin function from a careful combination of the results of
the POs $\Rb$, $\Ab$, and $\AFB$ \cite{ewpm} with the measurements of
the branching ratios of $\BXsga$ \cite{bsgamma} and $\BXsll$
\cite{bxsll}. In contrast to \cite{Bobeth:2005ck}, we do not include
the available experimental information on $\Kpnn$ \cite{kp} in our
global fit, as the constraining power of the latter measurement
depends in a non-negligible way on how the experimental ${\rm CL}$ of
the signal \cite{kpcl} is implemented in the analysis.

Third, and finally, we use the derived ranges for the Inami-Lim
functions in question to find lower and upper bounds for the branching
ratios of the rare decays $\Kpnn$, $\KLnn$, $\KLmm$, $\BXdsnn$, and
$\Bdsmm$ within CMFV.     

Our data set includes all POs measured at LEP and SLC that are related
to the $\Ztobb$ decay. It is given in \Tab{tab:pos}. Concerning the
used data we recall that the ratio $4/3 \, \AFB/\Al$ is lower than the
direct measurement of $\Ab$ by $1.6 \mysigma$, and lower than the SM
expectation for $\Ab$ by $3.2 \mysigma$ \cite{ewpm}. Whether this is
an experimental problem, an extreme statistical fluctuation or a real
NP effect in the bottom quark couplings is up to date
unresolved.\footnote{It has been known for some time that $\AFB$
  measured with respect to thrust axis is not infrared (IR) safe
  \cite{Catani:1999nf}. Recently, an IR safe definition of $\AFB$ has
  been suggested \cite{Weinzierl:2006yt} which defines the direction
  of the asymmetry by the jet axis after clustering the event with an
  IR safe flavor jet-algorithm \cite{Banfi:2006hf}. Given the
  long-standing discrepancy in $\AFB$ it would be interesting to
  reanalyze the existing data using this alternative definition.} In
fact, the relative experimental error in $\AFB$ is much larger than
the ones in the total $\Ztobb$ rate, $\Rb$, and $\Ab$, where no
anomalies are observed. Furthermore, the extracted value of the
anomalous LH coupling of the bottom quark agrees with its SM value
because of the strong constraint given by $\Rb$. This strong
constraint carries over to our results, which do not depend notably on
whether $\AFB$ is included in or excluded from the data set. We assume
that statistical fluctuations are responsible for the observed
discrepancy and include $\AFB$ in our global fit.

The actual calculations of $\Rb$, $\Ab$, and $\AFB$ used in our
analysis are performed with {\tt ZFITTER} \cite{zfitter}, which
includes the SM purely EW, QED and QCD radiative effects, photon
exchange and $\gamma$-$Z$ interference that are necessary to extract
the POs in a model-independent manner.\footnote{The default flags of
  {\tt ZFITTER} version 6.42 are used, except for setting ${\tt ALEM
    = 2}$ to take into account the externally supplied value of
  $\Daehad$.} For the purpose of our analysis, {\tt ZFITTER} has been
modified to include possible NP contributions to the $\Zbb$ vertex in
the parametrization of \Eq{eq:zdjdi}. The Higgs mass is allowed to
vary freely in the range $100 \, \GeV < \mh < 600 \, \GeV$. Since
$\Rb$ is largely insensitive to the mass of the Higgs boson this
conservative range has no noticeable impact on our results. The input
values of the other parameters entering $\Rb$, $\Ab$, and $\AFB$ are
collected in \App{app:input}.      

\begin{table}[!t]
\caption{\sf Results and correlations for the $\Ztobb$ POs of the fit
  to the LEP and SLC heavy flavor data taken from \cite{ewpm}.}   
\vspace{-3mm} 
\begin{center}
\begin{tabular}{c@{\hspace{2.5mm}}cccc}
\hline \hline \\[-4.5mm]
Observable & Result & \hspace{1mm} $\Rb$ & \hspace{1mm} $\Ab$ &
\hspace{1mm} $\AFB$ \\[0.5mm]
\hline 
$\Rb$ & $0.21629 \pm 0.00066$ & \hspace{1mm} $1.00$ & \hspace{1mm}
$-0.08$ & \hspace{1mm} $-0.10$ \\
$\Ab$ & $0.923 \pm 0.020$ & \hspace{1mm} & \hspace{1mm} $\phantom{+}
1.00$ & \hspace{1mm} $\phantom{+} 0.06$ \\  
$\AFB$ & $0.0992 \pm 0.0016$ & \hspace{1mm} & \hspace{1mm} &
\hspace{1mm} $\phantom{+} 1.00$ \\[1mm]
\hline \hline
\end{tabular}
\end{center}
\label{tab:pos}
\end{table}

The experimental results that we consider in connection with $\BXsga$
and $\BXsll$ are summarized in \Tab{tab:bs}. The given weighted
average of the branching ratio $\BRga$ corresponds to a photon energy
cut $E_\gamma > 1.6 \, \GeV$ in the $\bar{B}$-meson rest frame, while
for $\BRll$ the experimental data in the low-$q^2$ region $1 \, \GeV^2
< q^2 < 6 \, \GeV^2$ of the dilepton invariant mass squared, averaged
over electrons and muons are shown. Our calculations rely on
\cite{bsg} in the case of $\BXsga$ and on \cite{bsll} in the case of
$\BXsll$.  The used numerical input parameters can be found in
\App{app:input}. Unlike \cite{Bobeth:2005ck}, we do not include
$\BXsll$ data on the regions $0.04 \, \GeV^2 < q^2 < 1 \, \GeV^2$ and
$14.4 \, \GeV^2 < q^2 < 25 \, \GeV^2$ in our analysis. The reason for
this omission is twofold. First, in these regions the differential
$\BXsll$ rate is less sensitive to $\Delta C$ than in the low-$q^2$
region. Second, for high $q^2$ the theoretical uncertainties are
larger with respect to the ones that affect the low-$q^2$ region. An
inclusion of the latter two constraints would therefore make the fit
more complicated, but it would not improve the quality of the obtained
results.

\begin{table}[!t]
\caption{\sf World averages of $\BRga$ for $E_\gamma >
 1.6 \, \GeV$ and $\BRll$ for $1 \, \GeV^2 < q^2 < 6 \,\GeV^2$.}       
\vspace{-3mm} 
\begin{center}
\begin{tabular}{c@{\hspace{2.5mm}}c}
\hline \hline \\[-4.5mm]
Observable & \hspace{1mm} Result \\[0.5mm] 
\hline 
$\BRga \times 10^4$ & \hspace{1mm} $3.55 \pm 0.26$ \ \cite{Barbiero:2007cr} \\  
$\BRll \times 10^6$ & \hspace{1mm} $1.60 \pm 0.51$ \ \cite{bxsll} \\[1mm] 
\hline \hline
\end{tabular}
\end{center}
\label{tab:bs}
\end{table}

\begin{figure}[!t]
\vspace{-5mm}
\scalebox{0.4}{\includegraphics{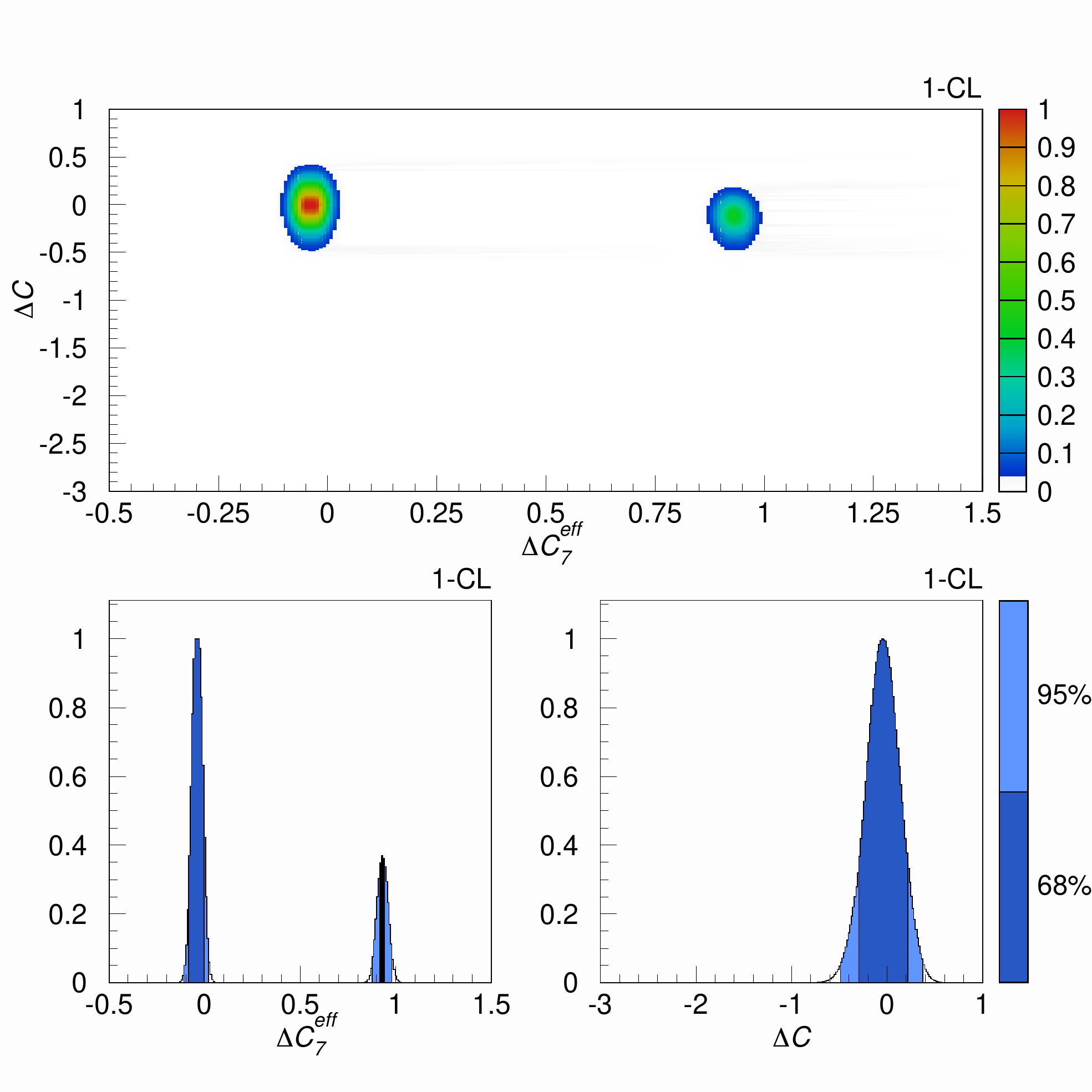}}

\vspace{-3mm}

\scalebox{0.4}{\includegraphics{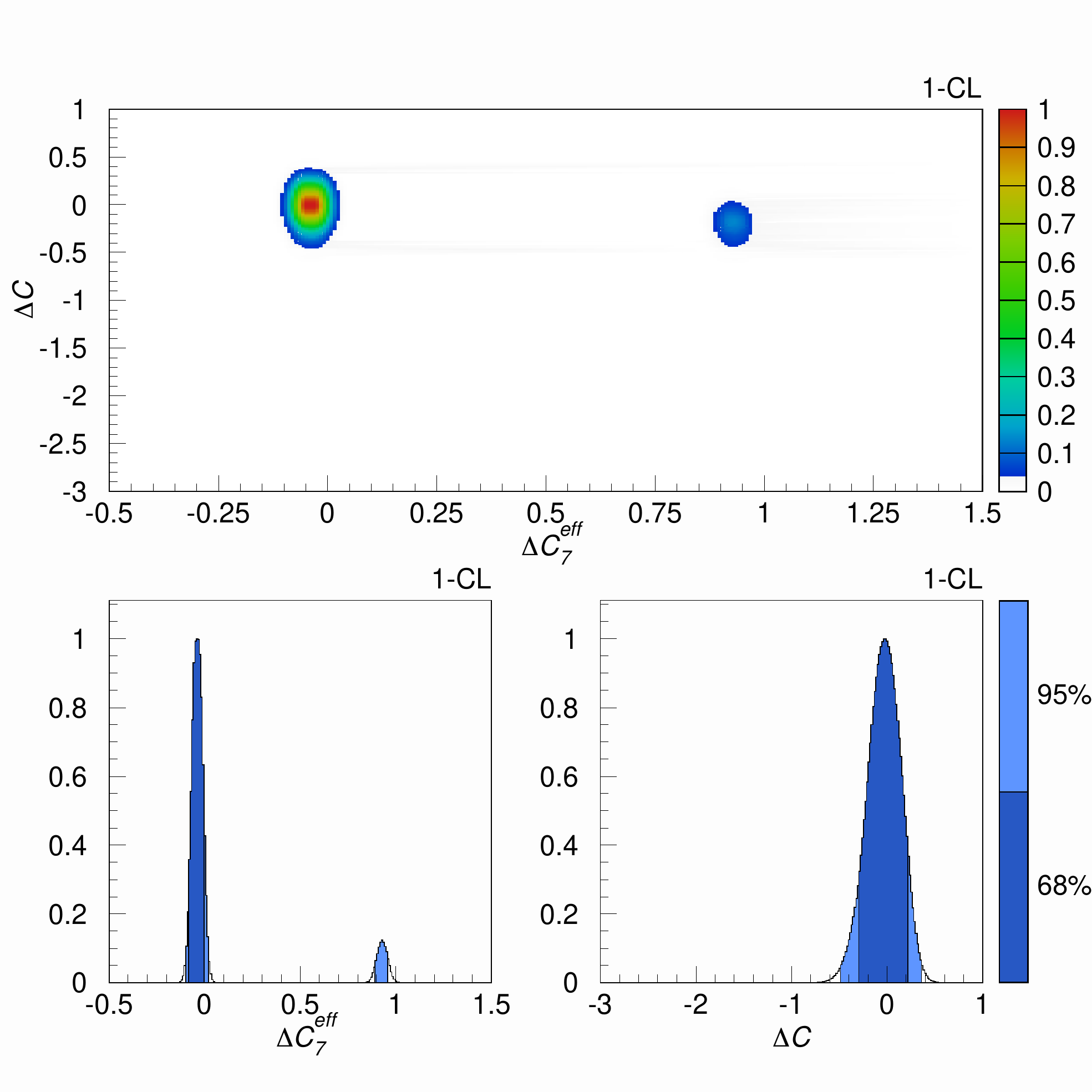}}
\vspace{-4mm}
\caption{\sf The upper (lower) panel displays the constraints on $\Delta
  C_7^{\rm eff}$ and $\Delta C$ within CMFV scanning $\Delta \Bll$ in
  the range $\pm 0.1$ (set to zero) that follow from a combination of
  the $\Ztobb$ POs with flavor observables. The colors encode the
  frequentist $1 - {\rm CL}$ level and the corresponding $68 \%$ and
  $95 \%$ probability regions as indicated by the bars on the right
  side of the panels. See text for details.}
\label{fig:dcdc7effnow}
\end{figure}

Before we present our final results, additional comments on the used
methodology concerning $\Delta \Bnn$, $\Delta \Bll$, $\Delta C$,
$\Delta D$, and $\Delta C_7^{\rm eff}$ are in order. We begin with
$\Delta C_7^{\rm eff}$ which enters both $\BXsga$ and $\BXsll$. A
well-known way to avoid the $\BXsga$ constraint consists in having a
large positive NP contribution $\Delta C_7^{\rm eff}$ that
approximately reverses the sign of the amplitude ${\cal A}
(\btosgamma) \propto C_7^{\rm eff} (\mb)$ with respect to the SM and
leaves $\BRga \propto |C_7^{\rm eff} (\mb)|^2$ unaltered within
experimental and theoretical uncertainties. In our analysis, we add
$\Delta C_7^{\rm eff}$ to the top quark contribution of the SM,
keeping $\mb$ that multiplies this combination renormalized at
$\mt$. This rescaling is motivated by the observation
\cite{Gambino:2001ew}, that in this way most of the logarithmic
enhanced QCD corrections are properly taken into account. We recall
that $C_{7 \, {\rm SM}}^{\rm eff} (\mb) \simeq -0.38$ in this
approach.

Both the value and the sign of $C_7^{\rm eff} (\mb)$ play an important
role in the $\BXsll$ decay rate \cite{wcbsll}. By contrast the
dependence of $\BRll$ on $\D$ is relatively weak. Nevertheless, for
suitable chosen values of $\Delta D$ the $\BXsll$ constraint can be
always satisfied even in the case of the non-SM solution of $\Delta
C_7^{\rm eff}$. In consequence, $\Delta D$ is not well constrained by
the data used, and we decided to scan $\Delta D$ in the range $\pm 1$
for the best fit value. This choice is rather generous since in the
CMFV scenarios that we consider one has $| \Delta D | < | D_{\rm SM}
|$ with $D_{\rm SM} \simeq -0.49$ throughout the allowed parameter
space \cite{Buras:2002ej, Blanke:2006eb, Buras:2000qz, dthdm}. We
verified that even larger variations have basically no effect on the
extraction of the allowed range for $\Delta C$, since the $\Ztobb$ POs
do not depend on $\Delta D$. The impact of $\Delta D$ on the bounds of
$\Delta C_7^{\rm eff}$ will be discussed below.

Precision data on $\Rb$, $\Ab$, and $\AFB$ lead to a tight, highly
model-independent constraint on $\re \, C_{\rm NP} (q^2 = \MZ^2)$. The
allowed range of $\Delta C$ can then be calculated from the identity
$\Delta C = (1 + \delta C_{\rm NP}) \re \, C_{\rm NP} (q^2 = \MZ^2)$
in any given model of NP where $\delta C_{\rm NP}$ is known. To carry
out the analysis in a generic way, one, however, needs to make an
assumption about the size of $\delta C_{\rm NP}$. Guided by the
results of \Secsand{sec:general}{sec:calculations} we allow $\delta
C_{\rm NP}$ to float in the range $\pm 0.1$. We note that larger
variations with, say, an absolute value of $| \delta C_{\rm SM} |
\simeq 0.3$, still lead to the conclusion that large negative values
of $\Delta C$ that would reverse the sign of $C_{\rm SM} \simeq 0.78$
are highly disfavored. 

The only EW box that enters the determination of $\Delta C_7^{\rm
  eff}$ and $\Delta C$ in our case is $\Delta \Bll$. To explore the
impact of the size of EW boxes on the fit results we consider two
scenarios. In the first we allow $\Delta \Bll$ to vary in the range
$\pm 0.1$ while in the second we assume $\Delta \Bll = 0$. The former
choice seems conservative, as relative to the SM value $\BllSM \simeq
0.18$ possible $\Delta \Bll$ contributions amount to only $^{+18}_{-5}
\%$ in the MSSM with MFV and small $\tan \beta$ \cite{Buras:2000qz}
and to below $+1 \%$ in both the mUED \cite{Buras:2002ej} and the LHT
model with degenerate mirror fermions \cite{Blanke:2006eb}. In fact,
the actual size of $\Delta \Bll$, which enters our fit through
$\BRll$, does only have a marginal effect on the results, because $\C$
is already tightly constraint by the combination of $\Rb$, $\Ab$, and
$\AFB$. Our bound on $\Delta C$ does not, for that reason, depend on
any conjecture concerning the size of EW boxes. Notice that this is
not the case in the analysis of the ``Magnificent Seven''
\cite{Bobeth:2005ck, M7}, which relies on the assumption $\Delta \Bnn
= \Delta \Bll = 0$ to derive a probability distribution function for
$\Delta C$.

The constraints on $\Delta C_7^{\rm eff}$ and $\Delta C$ within CMFV
following from the simultaneous use of $\Rb$, $\Ab$, $\AFB$, $\BRga$,
and $\BRll$ can be seen in \Fig{fig:dcdc7effnow}. All panels show
frequentist $1 - {\rm CL}$ levels. We see from the top and the lower
left plots that two regions, resembling the two possible signs of
$C_7^{\rm eff} (\mb)$, satisfy all existing experimental bounds. The
best fit value for $\Delta C_7^{\rm eff}$ is very close to the SM
point residing in the origin, while the wrong-sign solution located on
the right in the upper (lower) panel is barely (not) accessible at $68
\%$ probability, as it corresponds to a $\BRll$ value considerably
higher than the measurements \cite{Gambino:2004mv}. In the upper
(lower) panel of \Fig{fig:dcdc7effnow} the contribution $\Delta \Bll$
is scanned in the range $\pm 0.1$ (set to zero). In the former case
the full results read
\bea \label{eq:dc7effsb1}
\begin{gathered}
\Delta C_7^{\rm eff} = -0.039 \pm 0.043 \, \cup \, 0.931 \pm
0.016 \;\; (68 \% \, {\rm CL})
\, , \hspace{3mm} \\
\Delta C_7^{\rm eff} = [-0.104, 0.026] \, \cup \, [0.874, 0.988]
\;\; (95 \% \, {\rm CL}) \, , \hspace{2mm}
\end{gathered} 
\eea 
while in the latter one we obtain 
\bea \label{eq:dc7effsb0}
\begin{gathered}
\Delta C_7^{\rm eff} = -0.039 \pm 0.043 \;\; (68 \% \, {\rm CL})
\, , \hspace{2mm} \\ 
\Delta C_7^{\rm eff} = [-0.104, 0.026] \, \cup \, [0.890, 0.968] 
\;\;  (95 \% \, {\rm CL}) \, . \hspace{2mm}  
\end{gathered} 
\eea 
Similar bounds have been presented previously in
\cite{Bobeth:2005ck}. A comparison of \Eq{eq:dc7effsb1} with
\Eq{eq:dc7effsb0} makes clear that the size of $\Delta \Bll$ has only
a moderate impact on the accessibility of the non-SM solution of
$\Delta C_7^{\rm eff}$ while it leaves the ranges themselves almost
unchanged. Nevertheless, for $| \Delta \Bll | > | \Bll_{\rm SM} |$ the
wrong-sign case $\Delta C_7^{\rm eff} \simeq 0.93$ cannot be excluded
on the basis of $\BRga$ and $\BRll$ measurements alone. The same
statements apply to $\Delta D$ although its impact on the obtained
results is less pronounced than the one of $\Delta \Bll$. Notice that
since the SM prediction of $\BRga$ \cite{bsg} is now lower than the
experimental world average by more than $1 \mysigma$, extensions of
the SM that predict a suppression of the $\btosgamma$ amplitude are
strongly constrained. In particular, even the SM point $\Delta
C_7^{\rm eff} = 0$ is almost disfavored at $68 \% \, {\rm CL}$ by the
global fit. This tension is not yet significant, but could become
compelling once the experimental and/or theoretical error on $\BRga$
has been further reduced.

As can be seen from the top and the lower right plots in
\Fig{fig:dcdc7effnow}, in the case of $\Delta C$ only small
deviations from the SM are compatible with the data. In the upper
(lower) panel of \Fig{fig:dcdc7effnow} the contribution $\Delta \Bll$
is varied in the range $\pm 0.1$ (set to zero). In the former case we
find the following bounds
\beq \label{eq:dcsb1}
\begin{aligned}
\Delta C & = -0.037 \pm 0.266 & \!\! (68 \% \, {\rm CL}) \, , \\
\Delta C & = [-0.493, 0.387] & \!\! (95 \% \, {\rm CL}) \, ,
\end{aligned}
\eeq
while in the latter one we get 
\beq \label{eq:dcsb0}
\begin{aligned}
\Delta C & = -0.026 \pm 0.264 & \!\! (68 \% \, {\rm CL}) \, , \\
\Delta C & = [-0.483, 0.368] & \!\! (95 \% \, {\rm CL}) \, . 
\end{aligned}
\eeq
These results imply that large negative contributions to $\C$ that
would reverse the sign of the SM $Z$-penguin amplitude are highly
disfavored in CMFV scenarios due to the strong constraint from the
$\Ztobb$ POs, most notably, the one from $\Rb$. We stress that we
could not have come to this conclusion by considering only flavor
constraints, as done in \cite{Bobeth:2005ck}, since at present a
combination of $\BRga$, $\BRll$, and $\BRKp$ does not allow one to
distinguish the SM solution $\Delta C = 0$ from the wrong-sign case
$\Delta C \approx -2$. \Eqsand{eq:dcsb1}{eq:dcsb0} also show that the
derived bound on $\Delta C$ is largely insensitive to the size of
potential EW box contributions which is not the case if the $\Ztobb$
POs constraints are replaced by the one stemming from $\BRKp$.

\begin{figure}[!t]
\vspace{-5mm}
\scalebox{0.4}{\includegraphics{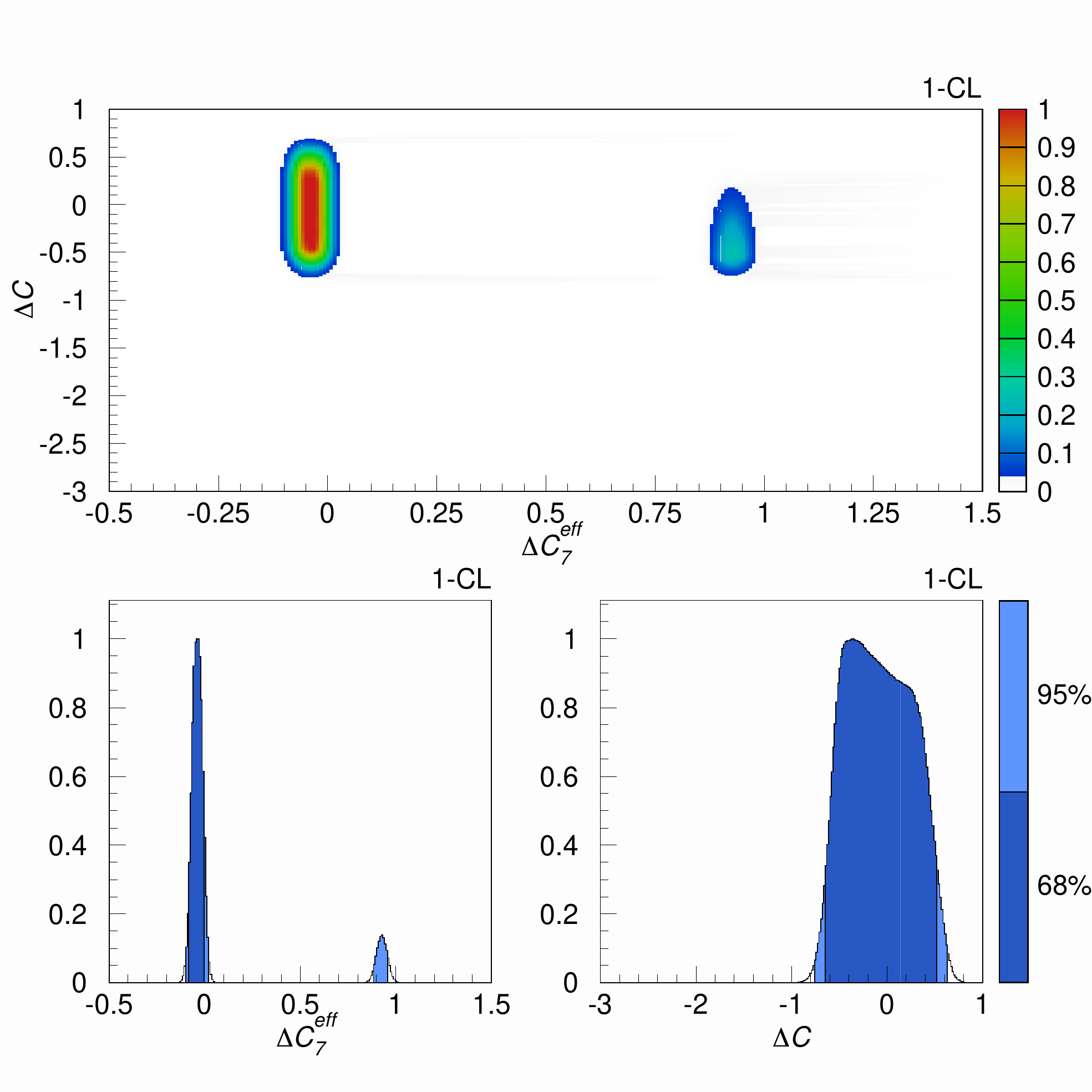}}

\vspace{-3mm}

\scalebox{0.4}{\includegraphics{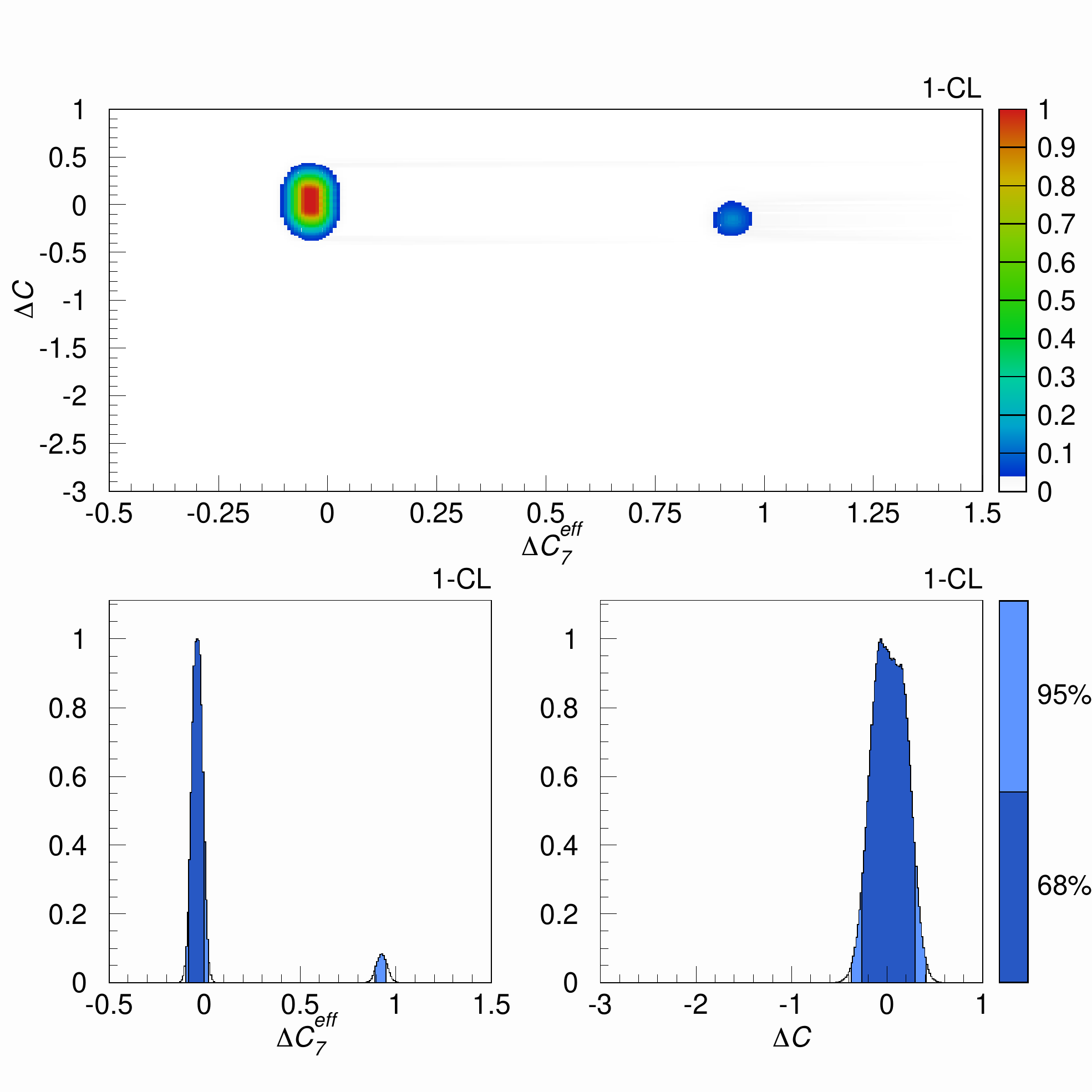}}
\vspace{-4mm}
\caption{\sf The upper (lower) panel displays future constraints on
  $\Delta C_7^{\rm eff}$ and $\Delta C$ within CMFV scanning $\Delta
  \Bnn$ in the range $\pm 0.4$ (set to zero) that are based on flavor
  observables only. The colors encode the frequentist $1 - {\rm CL}$
  level and the corresponding $68 \%$ and $95 \%$ probability regions
  as indicated by the bars on the right side of the panels. See text
  for details.}
\label{fig:dcdc7efffuture}
\end{figure}

It is easy to verify that the derived $95 \% \, {\rm CL}$ bound for
$\Delta C$ holds in each CMFV model discussed here. By explicit
calculations we find that the allowed range for $\Delta C$ is $[0,
0.12]$ and $[0, 0.13]$ in the THDM type I and II \cite{Bobeth:2001jm},
$[-0.05, 0.06]$ in the MSSM \cite{Buras:2000qz}, $[0, 0.04]$ in the
mUED scenario \cite{Buras:2002ej}, and $[-0.07, 0]$ in the LHT model
\cite{Blanke:2006eb} with degenerate mirror fermions.

Other theoretical clean observables that are sensitive to the
magnitude and sign of $\Delta C$ are the forward-backward and energy
asymmetries in inclusive and exclusive $\btosll$ decays \cite{wcbsll,
  Ali:1991is} and the branching ratios of $\KLnn$ and $\Kpnn$
\cite{Bobeth:2005ck, Buras:2004uu}. 

BaBar and Belle have recently reported measurements of the
forward-backward asymmetry $\AFBBKll$ \cite{Aubert:2006vb,
  fbabelle}. Both collaborations conclude that NP scenarios in which
the relative sign of the product of the effective Wilson coefficients
$C_9^{\rm eff} (\mb)$ and $C_{10}^{\rm eff} (\mb)$ is opposite to that
of the SM are disfavored at $95 \% \ {\rm CL}$. While these results
also point towards the exclusion of a large destructive NP
contribution to the $Z$-penguin amplitude, it is easy to verify that
the present $\AFBBKll$ constraint is less
restrictive\footnote{Assuming $\Delta \Bll = 0$, $\Delta D \lesssim
  4$, and neglecting all theoretical uncertainties leads to the very
  crude estimate $\Delta C \approx [-1, 1]$.} than the existing data
on $\Ztobb$ considered by us. Notice that a combination of the
branching ratios and forward-backward asymmetries of inclusive and
exclusive $\btosll$ transitions might in principle also allow one to
constrain the size of the NP contributions $\Delta \Bll$ and $\Delta
D$. A detailed study of the impact of all the available $\btosll$
measurements on the allowed range of $\Delta \Bll$, $ \Delta C$, and
$\Delta D$ is however beyond the scope of this article.

The remarkable power of the POs $\Rb$, $\Ab$, and $\AFB$ in unraveling
NP contributions to the universal Inami-Lim function $C$ is probably
best illustrated by a comparison to one of the undisputed
``heavyweight champions'' in this category, the $\Kpnn$ branching
ratio. A careful analysis of this decay shows that even under the
hypothesis $\Delta \Bnn = 0$ a future measurement of $\BRKp$ close to
its SM prediction with an accuracy of $\pm 10 \%$ would only lead to
a slightly better $95 \% \, {\rm CL}$ bound than the one given in
\Eq{eq:dcsb1}, while already relatively small deviations of $\Bnn$
from its SM value would give the $\Ztobb$ POs the edge. This feature
is illustrated in \Fig{fig:dcdc7efffuture} which shows the constraints
on $\Delta C$ and $\Delta C_7^{\rm eff}$ following from the
simultaneous use of the present $\BRga$ and $\BRll$ world averages and
a future accurate measurement of $\Kpnn$ leading to $\BRKp = (7.63 \pm
0.76) \times 10^{-11}$. In the upper (lower) panel the contribution
$\Delta \Bnn$ is allowed to float freely in the range $\pm 0.4$ (set
to zero). In the former case we find the following $\Delta C$ bounds
\beq \label{eq:Kdcsb1}
\begin{aligned}
\Delta C & = -0.057 \pm 0.588 & \!\! (68 \% \, {\rm CL}) \, , \\
\Delta C & = [-0.768, 0.668] & \!\! (95 \% \, {\rm CL}) \, ,
\end{aligned}
\eeq
while in the latter one we arrive at 
\beq \label{eq:Kdcsb0}
\begin{aligned}
\Delta C & = 0.026 \pm 0.282 & \!\! (68 \% \, {\rm CL}) \, , \\
\Delta C & = [-0.376, 0.420] & \!\! (95 \% \, {\rm CL}) \, . 
\end{aligned}
\eeq

However, this ``upset'' of the $\Kpnn$ mode should not be over
emphasized. While in MFV models the rates of the rare $\Knns$ decays
can be enhanced only moderately \cite{Bobeth:2005ck, Isidori:2006qy} a
very different picture can emerge in non-MFV scenarios with new
sources of flavor and $\CP$ violation. Since now the hard GIM
mechanism present in the MFV $\stodnn$ decay amplitude is in general
no longer active, large departures from the SM predictions are still
possible without violating any existing experimental constraint
\cite{Blanke:2006eb, kpnnnonmfv}. Precise measurements of the
processes $\Kpnn$ and $\KLnn$ will therefore have a non-trivial impact
on our understanding of the flavor structure and $\CP$ violation of NP
well above the $\TeV$ scale. This statement remains true even after
taking into account possible future constraints on the mass spectrum
obtained at the LHC and the refinement of the flavor constraints
expected from the $B$-factories \cite{Isidori:2006qy, Blanke:2006eb}.

\begin{widetext}
\begin{center}
\begin{table}[!t]
\caption{\sf Bounds for various rare decays in CMFV models at $95 \%$
  probability, the corresponding values in the SM at $68 \%$ and $95
  \% \, {\rm CL}$, and the available experimental information. See
  text for details.}    
\vspace{1mm} 
\begin{center}
\begin{tabular}{c@{\hspace{2.5mm}}cccc}
\hline \hline \\[-4.5mm]
Observable & CMFV ($95 \% \, {\rm CL}$) & \hspace{0mm} SM ($68 \% \,
{\rm CL}$) & \hspace{0mm} SM ($95 \% \, {\rm CL}$) & \hspace{0mm}
Experiment \\[0.5mm] 
\hline 
$\BRKp \times 10^{11}$ & $[4.29, 10.72]$ & \hspace{0mm} $7.15 \pm 1.28$ &  
\hspace{0mm} $[5.40, 9.11]$ & \hspace{0mm} $\left ( 14.7^{+13.0}_{-8.9}
\right )$ \cite{kp} \\
$\BRKL \times 10^{11}$ & $[1.55, 4.38]$ & \hspace{0mm} $2.79 \pm 0.31$
& \hspace{0mm} $[2.21, 3.45]$ & \hspace{0mm} $< 2.1 \times 10^4 \;\;
(90 \% \, \text{CL})$ \cite{Ahn:2006uf} \\
$\BRKm \times 10^9$ & $[0.30, 1.22]$ & \hspace{0mm} $0.70 \pm 0.11$ &
\hspace{0mm} $[0.54, 0.88]$ & \hspace{0mm} -- \\
$\BRXd \times 10^6$ & $[0.77, 2.00]$ & \hspace{0mm} $1.34 \pm 0.05$ &
\hspace{0mm} $[1.24, 1.45]$ & \hspace{0mm} -- \\
$\BRXs \times 10^5$ & $[1.88, 4.86]$ & \hspace{0mm} $3.27 \pm 0.11$ &
\hspace{0mm} $[3.06, 3.48]$ & \hspace{0mm} $< 64 \;\; (90 \% \,
\text{CL})$ \cite{Barate:2000rc} \\ 
$\BRBd \times 10^{10}$ & $[0.36, 2.03]$ & \hspace{0mm} $1.06 \pm 0.16$ &
\hspace{0mm} $[0.87, 1.27]$ & \hspace{0mm} $< 3.0 \times 10^2 \;\; (95
\% \, \text{CL})$ \cite{Bernhard:2006fa} \\  
$\BRBs \times 10^9$ & $[1.17, 6.67]$ & \hspace{0mm} $3.51 \pm 0.50$ &
\hspace{0mm} $[2.92, 4.13]$ & \hspace{0mm} $< 9.3 \times 10^1
\;\; (95 \% \, \text{CL})$ \cite{Sanchez:2007ew} \\[1mm] 
\hline \hline
\end{tabular}
\end{center}
\label{tab:brs}
\end{table}
\end{center}
\end{widetext}

In order to allow a better comparison with the results presented
previously in \cite{Bobeth:2005ck}, we will set $\Delta \Bnn = \Delta
\Bll = 0$ when determining the allowed ranges for the branching ratios
of $\Kpnn$, $\KLnn$, $\KLmm$, $\BXdsnn$, and $\Bdsmm$ within CMFV. The
corresponding lower and upper bounds at $95 \%$ probability are
reported in \Tab{tab:brs}. For comparison, we also show the $68 \%$
and $95 \% \ {\rm CL}$ limits in the SM, obtained using the CKM
parameters from a standard UT analysis. The calculations of the SM
branching ratios all employ the results of \cite{XY}. In addition we
take into account the recent theoretical developments of
\cite{Isidori:2005xm, bghn, Mescia:2007kn} in the case of $\Kpnn$ and
$\KLnn$, and of \cite{Gorbahn:2006bm} for what concerns $\KLmm$. In
contrast to the standard approach we normalize the $\BXdsnn$ decay
width to the $\BXuen$ rate, while we follow \cite{Buras:2003td} in the
case of $\Bdsmm$, since both procedures lead to a reduction of
theoretical uncertainties. The actual numerical analysis is performed
with a modified version of the CKMfitter code. The used input
parameters are given in \App{app:input}.

It is evident from \Tab{tab:brs} that the strong bound on $\Delta C$,
coming mainly from the existing precision measurements of the $\Ztobb$
POs, does only allow for CMFV departures relative to the SM branching
ratios that range from around $\pm 20 \%$ to at most $\pm 60 \%$ for
the given rare $K$- and $B$-decays. While our upper bounds are in
good agreement with the results of \cite{Bobeth:2005ck}, the derived
lower bounds are one of the new results of our article. A strong
violation of any of the $95 \% \ {\rm CL}$ bounds on the considered
branching ratios by future measurements will imply a failure of the
CMFV assumption, signaling either the presence of new effective
operators and/or new flavor and $\CP$ violation. A way to evade the
given limits is the presence of sizable corrections $\delta C_{\rm
  NP}$ and/or $\Delta \Bnn$ and $\Delta \Bll$. While these
possibilities cannot be fully excluded, general arguments and explicit
calculations indicate that they are both difficult to realize in the
CMFV framework.

\section{Conclusions}
\label{sec:conclusions}

To conclude, we have pointed out that large contributions to the
universal Inami-Lim function $\C$ in constrained
minimal-flavor-violation that would reverse the sign of the standard
$Z$-penguin amplitude are highly disfavored by the existing
measurements of the pseudo observables $\Rb$, $\Ab$, and $\AFB$
performed at LEP and SLC. This underscores the outstanding role of
electroweak precision tests in guiding us toward the right theory and
immediately raises the question: What else can flavor physics learn
from the high-energy frontier?

\acknowledgments{We thank A.~J.~Buras and C.~Tarantino for their
  careful reading of the manuscript and for advertising our work
  before publication. We are indebted to M.~Blanke, C.~Bobeth,
  T.~Ewerth, A.~Freitas, T.~Hahn, A.~H\"ocker, T.~Huber, M.~Misiak,
  J.~Ocariz, A.~Poschenrieder, C.~Smith, and M.~Spranger for helpful
  discussions and correspondence. Color consultation services provided
  by A.~Daleo and G.~Zanderighi are acknowledged. This work has been
  supported in part by the Schweizer Nationalfonds and the National
  Science Foundation under Grant PHY-0355005. Our calculations made
  use of the zBox1 computer at the Universit\"at Z\"urich
  \cite{zbox}.}

\appendix

\section{Vertex functions}
\label{app:cfunctions}

Below we present the analytic expressions for the non-universal
contributions to the renormalized LH $\Zdjdi$ vertex functions in the
CMFV models considered in this article. All expressions correspond to
the limit of on-shell external quarks with vanishing mass and have
been obtained in the 't Hooft-Feynman gauge.

In the THDM type I, the only additional non-universal contribution to
the $\Ztodjdi$ transitions stems from diagrams with charged Higgs
boson, $H^\pm$, and top quark, $t$, exchange. An example of such a
graph is shown on the top left-hand side of \Fig{fig:cmfv}. The
correction depends on the mass of the charged Higgs boson, $\MHpm$,
the top quark mass, $\mt$, and on the ratio of the vacuum expectation
value of the Higgs doublets, $\tan \beta$. Using the decomposition
of \Eq{eq:zdjdi} we find for the corresponding form factor
\begin{widetext}
\begin{align} \label{eq:CTHDM}
  C_{\rm THDM} (q^2) & = \f{\tan^2 \beta}{96} \f{\mt^2}{\MW^2}
  \scalebox{1}{\Bigg [} \f{(22 \sw^2-9) \MHpms - 3 \, (6 \sw^2-1)
    \mt^2}{\MHpms - \mt^2}
  \non \\
  & + \f{2 \, (2 \sw^2-3) \left (2 \MHpmq - (4 \mt^2 + q^2) \MHpms + 2
      (\mt^2 + q^2) \mt^2 \right)}{(\MHpms - \mt^2)^2 q^2} A_0
  (\MHpms)
  \non \\
  & - \f{2 \, (2 \sw^2-3) \left (2 \MHpmq - 4 \MHpms \mt^2 + 2 \mt^4 +
      \mt^2 q^2 \right)}{(\MHpms - \mt^2)^2 q^2} A_0(\mt^2)
  \non \\
  & - \f{6 \, (2 \sw^2-1) \left (2 \MHpms - 2 \mt^2 - q^2
    \right)}{q^2}
  B_0(q^2,\MHpms,\MHpms) \non \\
  & + \f{8 \sw^2 \left (2 \MHpms - 2 \mt^2 - q^2 \right)}{q^2}
  B_0(q^2,\mt^2,\mt^2) \non \\
  & + \f{12 \, (2\sw^2-1) \left (\MHpmq - 2 \MHpms \mt^2 + \mt^4 +
      \mt^2 q^2 \right)}{q^2} \hspace{0.5mm}
  C_0(q^2,0,0,\MHpms,\MHpms,\mt^2) \non \\
  & + \f{4 \left (4 \sw^2 \MHpmq + 4 \sw^2 \mt^4 - \left ( 8 \sw^2
        \MHpms - (4 \sw^2-3) q^2 \right) \mt^2 \right)}{q^2}
  \hspace{0.5mm} C_0(q^2,0,0,\mt^2,\mt^2,\MHpms)
  \scalebox{1}{\Bigg ]} \, . 
\end{align}
\end{widetext}
Here and in the following the coefficients $A_0 (m^2)$, $B_0
(p^2,m_1^2,m_2^2)$, and $C_0
(p_1^2,p_2^2,(p_1+p_2)^2,m_1^2,m_2^2,m_3^2)$ denote the finite parts
of the scalar one-, two-, and three-point functions in the $\MSbar$
scheme as implemented in {\it LoopTools} \cite{Hahn:1998yk} and {\it
  FF} \cite{vanOldenborgh:1990yc}. The above result agrees with the
findings of \cite{Denner:1991ie}. In the case of the THDM type II the
overall factor $\tan^2 \beta$ has to be replaced by $\cot^2 \beta$.

In the case of the MSSM only SUSY diagrams involving chargino,
$\tilde{\chi}^\pm_m$, and up-squark, $\tilde{u}_a$, exchange lead to
non-universal correction to the renormalized LH $\Zdjdi$ vertex of
\Eq{eq:zdjdi}. An example of a possible contribution can be seen on
the top right side of \Fig{fig:cmfv}. The corresponding form factor
can be written as
\begin{widetext}
\begin{align} \label{eq:CCha}
   C_{\tilde{\chi}^\pm} (q^2) & = -\f{\kappa_{ij} e^2}{4 \MW^2}
  \sum_{m, n = 1}^2 \sum_{a, b = 1}^6
  (X^{U_{L}\dagger}_m)_{jb}(X^{U_{L}}_n)_{ai} \scalebox{1.185}{\Bigg
    \{} 2 M_{\tilde{\chi}_n}^{\pm} M_{\tilde{\chi}_m}^{\pm} C_0
  (q^2,0, 0, M_{\tilde{\chi}_n}^{\pm 2}, M_{\tilde{\chi}_m}^{\pm 2},
  m_{\tilde{u}_a}^2) \hspace{0.5mm}
  U_{m1} U_{n1}^\ast \delta_{ab} \non \\
  & + \Bigg [ -\ln (M_{\tilde{\chi}_n}^{\pm 2}) + \f{1}{2 q^2} \bigg (
  3 q^2 + 2 A_0 (M_{\tilde{\chi}_n}^{\pm 2}) + 2 A_0
  (M_{\tilde{\chi}_m}^{\pm 2})
  - 4 A_0 (m_{\tilde{u}_a}^2) \non \\
  & \hspace{2em} - 2 \left ( M_{\tilde{\chi}_n}^{\pm 2} +
    M_{\tilde{\chi}_m}^{\pm 2} - 2 m_{\tilde{u}_a}^2 + q^2 \right )
  B_0
  (q^2, M_{\tilde{\chi}_n}^{\pm 2}, M_{\tilde{\chi}_m}^{\pm 2}) \non \\
  & \hspace{2em} + 4 \hspace{0.5mm} (M_{\tilde{\chi}_n}^{\pm 2} -
  m_{\tilde{u}_a}^2) ( M_{\tilde{\chi}_m}^{\pm 2} - m_{\tilde{u}_a}^2
  ) \hspace{0.5mm} C_0 (q^2, 0, 0, M_{\tilde{\chi}_n}^{\pm 2},
  M_{\tilde{\chi}_m}^{\pm 2},
  m_{\tilde{u}_a}^2) \bigg ) \Bigg ] W_{m1}^\ast W_{n1} \delta_{ab} \non \\
  & - \Bigg [ -\ln (M_{\tilde{\chi}_n}^{\pm 2}) + \f{1}{2 q^2} \bigg (
  q^2 + 4 A_0 (M_{\tilde{\chi}_n}^{\pm 2}) - 2 A_0 (m_{\tilde{u}_a}^2)
  - 2 A_0 (m_{\tilde{u}_b}^2) \non \\
  & \hspace{2em} - 2 \left ( 2 M_{\tilde{\chi}_n}^{\pm 2} -
    m_{\tilde{u}_a}^2 - m_{\tilde{u}_b}^2 + q^2 \right ) B_0
  (q^2, m_{\tilde{u}_a}^2, m_{\tilde{u}_b}^2) \non \\
  & \hspace{2em} - 4 \left ( M_{\tilde{\chi}_n}^{\pm 4} +
    m_{\tilde{u}_a}^2 m_{\tilde{u}_b}^2 - \left ( m_{\tilde{u}_a}^2 +
      m_{\tilde{u}_b}^2 - q^2 \right ) M_{\tilde{\chi}_n}^{\pm 2}
  \right ) \hspace{0.5mm} C_0 (q^2, 0, 0, m_{\tilde{u}_a}^2,
  m_{\tilde{u}_b}^2, M_{\tilde{\chi}_n}^{\pm 2} ) \bigg ) \Bigg ]
  (\Gamma^{U_{L}}
  {\Gamma^{U_{L}\dagger}})_{b a} \delta_{mn} \non \hspace{7.5mm} \\
  & + \Bigg [ \f{-(22 \sw^2 - 9) \hspace{0.5mm}
    M_{\tilde{\chi}_n}^{\pm 2} + 3 \hspace{0.5mm} (6 \sw^2 - 1)
    \hspace{0.5mm} m_{\tilde{u}_a}^2}{6
    (M_{\tilde{\chi}_n}^{\pm 2} - m_{\tilde{u}_a}^2)} \non \\
  & \hspace{2em} - \f{(2 \sw^2 - 3) \left ( 2 M_{\tilde{\chi}_n}^{\pm
        4} + 2 m_{\tilde{u}_a}^4 - (4 m_{\tilde{u}_a}^2 - q^2)
      \hspace{0.5mm} M_{\tilde{\chi}_n}^{\pm 2} \right )}{3
    (M_{\tilde{\chi}_n}^{\pm 2} - m_{\tilde{u}_a}^2)^2 q^2} A_0
  (M_{\tilde{\chi}_n}^{\pm 2}) \non \\
  & \hspace{2em} + \f{(2 \sw^2 - 3) \left ( 2 M_{\tilde{\chi}_n}^{\pm
        4} + 2 m_{\tilde{u}_a}^4 - 2 \, (2 m_{\tilde{u}_a}^2 - q^2)
      \hspace{0.5mm} M_{\tilde{\chi}_n}^{\pm 2} - m_{\tilde{u}_a}^2
      q^2\right )}{3 (M_{\tilde{\chi}_n}^{\pm 2} -
    m_{\tilde{u}_a}^2)^2 q^2}
  A_0 (m_{\tilde{u}_a}^2) \non \\
  & \hspace{2em} + \f{(2 \sw^2 - 1) \left ( 2 M_{\tilde{\chi}_n}^{\pm
        2} - 2 m_{\tilde{u}_a}^2 + q^2 \right )}{q^2} B_0 (q^2,
  M_{\tilde{\chi}_n}^{\pm 2}, M_{\tilde{\chi}_n}^{\pm 2}) \non \\
  & \hspace{2em} - \f{4 \sw^2 \left ( 2 M_{\tilde{\chi}_n}^{\pm 2} - 2
      m_{\tilde{u}_a}^2 + q^2 \right )}{3 q^2}
  B_0 (q^2, m_{\tilde{u}_a}^2, m_{\tilde{u}_a}^2) \non \\
  & \hspace{2em} - \f{2 \, (2 \sw^2 - 1) \left (
      M_{\tilde{\chi}_n}^{\pm 4} + m_{\tilde{u}_a}^4 - (2
      m_{\tilde{u}_a}^2 - q^2) \hspace{0.5mm} M_{\tilde{\chi}_n}^2
    \right )}{q^2} C_0 (q^2, 0, 0,
  M_{\tilde{\chi}_n}^{\pm 2}, M_{\tilde{\chi}_n}^{\pm 2},
  m_{\tilde{u}_a}^2) \non \\ 
  & \hspace{2em} - \f{8 \sw^2 \left ( M_{\tilde{\chi}_n}^{\pm 4} +
      m_{\tilde{u}_a}^4 - (2 m_{\tilde{u}_a}^2 - q^2) \hspace{0.5mm}
      M_{\tilde{\chi}_n}^{\pm 2} \right )}{3 q^2} C_0 (q^2, 0, 0,
  m_{\tilde{u}_a}^2, m_{\tilde{u}_a}^2, M_{\tilde{\chi}_n}^{\pm 2})
  \Bigg ] \delta_{ab} \delta_{mn} \scalebox{1.185}{\Bigg \}} \, , 
\end{align}
\end{widetext}
where $\kappa_{ij} \equiv (8 \sqrt{2} \GF e^2 V^\ast_{tj}
V_{ti})^{-1}$ with $V$ being the CKM matrix and $i, j = d, s, b$.

The LH chargino-up-squark-down-quark coupling matrix takes the form  
\bea \label{eq:XnUL}
X_n^{U_L} = -\f{e}{\sw} \left ( W_{n1}^{\ast} \Gamma^{U_L} -
  W_{n2}^{\ast} \Gamma^{U_R} \f{M_U}{\sqrt{2} \MW s_\beta} \right) V
\, . \hspace{2.5mm}
\eea
Here and in the following $s_\beta \equiv \sin \beta$, $c_\beta \equiv
\cos \beta$, $t_\beta \equiv \tan \beta$, etc. 

The unitary mixing matrices $U$ and $W$ are defined through 
\beq \label{eq:UVdef}
U^{\ast} M_{\tilde{\chi}^\pm} W^{\dagger} = {\rm diag}
(M_{\tilde{\chi}_1}^{\pm}, M_{\tilde{\chi}_2}^{\pm}) \, ,
\eeq
with $M_{\tilde{\chi}_{1,2}}^{\pm}$ being the physical chargino masses
that satisfy $M^{\pm}_{\tilde{\chi}_1} < M^{\pm}_{\tilde{\chi}_2}$.
$M_{\tilde{\chi}^\pm}$ denotes the chargino mass matrix, which in
terms of the wino, $M_2$, and higgsino mass parameter, $\mu$, reads
\beq \label{eq:MCha}
M_{\tilde{\chi}^\pm} =
\begin{pmatrix}
M_2 & \sqrt{2} \MW s_\beta \\ 
\sqrt{2} \MW c_\beta & \mu 
\end{pmatrix} \, .
\eeq

The $6\times 3$ matrices 
\beq \label{eq:GammaULR}
(\Gamma^{U_L})_{ai} = (\Gamma^U)_{ai} \, , \hspace{5mm} 
(\Gamma^{U_R})_{ai} = (\Gamma^U)_{a, i+3} \, .
\eeq
are building blocks of the unitary matrix $\Gamma^U$ that diagonalizes
the $6 \times 6$ mass-squared matrix $M^2_{\tilde U}$ of the up-type
squarks:
\beq
\Gamma^U M_{\tilde U}^2 {\Gamma^U}^{\dagger}= 
{\rm diag} (m_{\tilde{u}_1}^2, \dots, m_{\tilde{u}_6}^2) \, .
\eeq
In the super-CKM basis \cite{Misiak:1997ei}, $M^2_{\tilde U}$ is given
by
\begin{widetext}
\beq \label{eq:Msup}
M^2_{\tilde U} =
\begin{pmatrix}
  M_{\tilde{U}_L}^2 + M_U^2 + \MZ^2 \hspace{0.5mm} c_{2 \beta} \left (
    \f{1}{2} - \f{2}{3} \sws \right ) \unit
  & M_U \left ( A_{U}^\ast - \mu \hspace{0.5mm} t_\beta^{-1} \unit \right ) \\
  \left [ M_U \left ( A_{U}^\ast - \mu \hspace{0.5mm} t_\beta^{-1}
      \unit \right ) \right ]^\dagger & M_{\tilde{U}_R}^2 + M_U^2 +
  \frac{2}{3} \MZ^2 \hspace{0.5mm} c_{2 \beta} \sws \unit
\end{pmatrix} \, ,
\eeq
\end{widetext}
where $M_{\tilde{U}_L} = m_{\tilde{Q}_L} \unit$ and $M_{\tilde{U}_R} =
m_{\tilde{u}_R} \unit$ are the left and right soft SUSY breaking
up-type squark mass matrices, $M_U = {\rm diag} (m_u, m_c, m_t)$, $A_U
= A_u \unit$ contains the trilinear parameters and $\unit$ represents
the $3 \times 3$ unit matrix. We assume $\CP$ conservation, so all
soft SUSY breaking terms are real.

The result given in \Eq{eq:CCha} agrees with the one of
\cite{Boulware:1991vp}. To verify the consistency of the results one
has to take into account that in the last equation of
\cite{Boulware:1991vp} the coefficient $C_{11}$ should read $C_{12}$,
and that arbitrary constant terms can be added to the second and third
coefficient of the four different coupling structures in \Eq{eq:CCha},
since their contribution disappears after the summations over $m, n ,
a$, and $b$ have been performed. In addition, the explicit $\ln
(M_{\tilde{\chi}_n}^{\pm 2})$ terms are absent in
\cite{Boulware:1991vp}. They have been chosen such that
$C_{\tilde{\chi}^\pm} (q^2)$ coincides for $q^2 = 0$ with the
expression for the one-loop $Z$-penguin function given in
\cite{Bobeth:2001jm}.

In the mUED model diagrams containing infinite towers of the KK modes
corresponding to the $W$-boson, $W^\pm_{(k)}$, the pseudo Goldstone
boson, $G^\pm_{(k)}$, the $SU (2)$ quark doublets, ${\cal Q}_{q (k)}$,
and the $SU (2)$ quark singlets, ${\cal U}_{q (k)}$, as well as the
charged scalar, $a^\pm_{(k)}$, contribute to the non-universal
correction to the $\Zdjdi$ vertex. A possible diagram can be seen on
the lower left side in \Fig{fig:cmfv}. The only additional parameter
entering the form factor in \Eq{eq:zdjdi} relative to the SM is the
inverse of the compactification radius $1/R$. We obtain
\begin{widetext}
\begin{align} \label{eq:CACD}
  C_{\rm mUED} (q^2) & = \f{1}{96} \sum^{\infty}_{k = 1}
  \scalebox{1.185}{\Bigg [} \f{\left(-8 \, (2 \sw^2-3) \mks - (34
      \sw^2 - 27) \MW^2 + 3 \, (6 \sw^2-1) \mt^2 \right) \mt^2}{(\mt^2
    - \MW^2) \MW^2} \non \\
  & - \f{2 \, (2 \sw^2-3)}{(\mt^2 - \MW^2)^2 \MW^2 q^2} \Big (2 \mt^6
  + (2 \MW^2 + q^2) \mt^4 - 5 \,
  (2 \MW^2 + q^2) \hspace{0.5mm} \mt^2 \MW^2 \non \\
  & \hspace{5em} + 6 \MW^6 + (\mt^2 +
  3 \MW^2) \hspace{0.5mm} \mks q^2 + 8 \MW^4 q^2 \Big )  A_0(\mtks) \non \\
  & + \f{2 \, (2 \sw^2-3) \left ( \left (2 \mt^4 - (4 \MW^2 + 3 q^2)
        \mt^2 + 2 \MW^4 + 7 \MW^2 q^2 \right) \MW^2 + (7 \MW^2 -3
      \mt^2) \hspace{0.5mm} \mks q^2 \right) \mt^2}{ (\mt^2 - \MW^2)^2
    \MW^4 q^2} A_0(\MWks) \non \\
  & + \f{2 \, (2 \sw^2-3) \left (6 \MW^4 + 3 \mks q^2 + 8 \MW^2 q^2
    \right)}{\MW^4 q^2} A_0(\mks) \non \\
  & - \f{2 \left(8 \sw^2 \mt^4 + 2 \left( (8 \sw^2-9) \MW^2 + 2 \sw^2
        q^2 \right) \mt^2 - (4 \sw^2-3) (6 \MW^2 + 5 q^2) \MW^2
    \right) }{\MW^2 q^2} B_0(q^2,\mtks,\mtks) \non \\
  & + \f{6 \left((4 \sw^2-2) \hspace{0.5mm} \mt^2 + 2 (4 \sw^2-5)
      \MW^2 + (2 \sw^2-1) \hspace{0.5mm} q^2 \right) \mt^2}{\MW^2 q^2}
  B_0(q^2,\MWks,\MWks) \non \\
  & - \f{2 \, (4 \sw^2-3) (6 \MW^2 + 5 q^2)
  }{q^2} B_0(q^2,\mks,\mks) \\
  & + \f{4}{\MW^2 q^2} \Big( 4 \sw^2 \mt^6 + \left ((4
    \sw^2-9) \MW^2 + (4 \sw^2-3) \hspace{0.5mm} q^2 \right) \mt^4 \non \\
  & \hspace{5em} - 2 \left ((10 \sw^2-9) \MW^4 - 2 \sw^2 \mks q^2 + 2
    \, (\sw^2-3) \MW^2 q^2 \right) \mt^2 \non \\
  & \hspace{5em} + (4 \sw^2-3) \left( 3 \MW^4 + 3 \mks q^2 + 4 \MW^2
    q^2 + 2 q^4
  \right) \MW^2 \Big) C_0(q^2,0,0,\mtks,\mtks,\MWks) \non \\
  & + \f{12}{\MW^2q^2} \Big ((2 \sw^2-1) \mt^6 + \left (2 \,
    (\sw^2-2) \MW^2 + (2 \sw^2-1) \hspace{0.5mm} q^2 \right) \mt^4 \non \\
  & \hspace{5em} - \left ((10 \sw^2-11) \MW^4 - (2 \sw^2-1) \mks q^2
    + 2 \, (\sw^2+1) \MW^2 q^2 \right) \mt^2 \non \\
  & \hspace{5em} - 2 \cw^2 \left (3 \MW^4 + 3 \mks q^2 + 4 \MW^2 q^2
  \right) \MW^2 \Big ) \hspace{0.5mm} C_0(q^2,0,0,\MWks,\MWks,\mtks) \non \\
  & + \f{24 \, \cw^2 \left(3 \MW^4 + 3 \mks q^2 + 4 \MW^2 q^2
    \right)}{q^2} \hspace{0.5mm} C_0(q^2,0,0,\MWks,\MWks,\mks) \non \\
  & - \f{4 \, (4 \sw^2-3) \left( 3 \MW^4 + 3 \mks q^2 + 4 \MW^2 q^2 +
      2 q^4 \right)}{q^2} \hspace{0.5mm} C_0(q^2,0,0,\mks,\mks,\MWks)
  \scalebox{1.185}{\Bigg ]} \, , \non 
\end{align}
\end{widetext}
where $\mtk = \sqrt{\mt^2 + \mks}$, $\MWk = \sqrt{\MW^2 + \mks}$, and
$\mk = k/R$. We note that our new result for $C_{\rm mUED} (q^2)$
coincides for $q^2 = 0$ with the one-loop KK contribution to the
$Z$-penguin function found in \cite{Buras:2002ej}.

In the case of the LHT with degenerate mirror fermions the only new
particle that effects the $\Ztodjdi$ transition in a non-universal way
is a $T$-even heavy top, $T_+$. A sample diagram involving such a
fermion is given on the lower right-hand side of \Fig{fig:cmfv}. The
form factor depends on the mass of the heavy top, which is
controlled by the size of the top Yukawa coupling, the symmetry
breaking scale $f$, and the parameter $x_L \equiv
\lambda_1^2/(\lambda_1^2 + \lambda_2^2)$. Here $\lambda_1$ is the
Yukawa coupling between $t$ and $T_+$ and $\lambda_2$ parametrizes the
mass term of $T_+$. Our result for the form factor entering
\Eq{eq:zdjdi} is given by
\begin{widetext}
\begin{align} \label{eq:CLHT}
  C_{\rm LHT} (q^2) & = \f{x^2_L}{96} \f{v^2}{f^2}
  \scalebox{1.185}{\Bigg [} \f{3 (\MTps - \mt^2) \left ( \left ((6
        \sw^2-1) \MTps - (10 \sw^2-7) \MW^2 \right) \MW^2 - (6
      \sw^2-1) (\MTps - \MW^2) \hspace{0.5mm}
      \mt^2 \right)}{(\MTps - \MW^2) (\mt^2 - \MW^2) \MW^2} \non \\
  & + \f{2 \, (2 \sw^2-3) \left (2 \MTpt + \MTpq q^2 - 6 (\MW^2 + q^2)
      \MTps \MW^2 + 4 (\MW^2 + 2 q^2) \MW^4 \right)}{(\MTps - \MW^2)^2
    \MW^2 q^2} A_0(\MTps) \non \\
  & - \f{2 \, (2 \sw^2-3) \left (2 \mt^6 + \mt^4 q^2 - 6 (\MW^2 + q^2)
      \hspace{0.5mm} \mt^2 \MW^2 + 4 (\MW^2 + 2 q^2) \MW^4
    \right)}{(\mt^2 - \MW^2)^2 \MW^2 q^2} A_0(\mt^2) \non \\
  & + \f{2 \, (2 \sw^2-3) (\mt^2 - \MTps)}{(\MTps - \MW^2)^2 (\mt^2 -
    \MW^2)^2 \MW^2 q^2} \Big (2 (\MTps - \MW^2)^2 \mt^4 \non \\
  & \hspace{5em} - \left (4 \MTpq - (8 \MW^2 + q^2) \MTps + 4 (\MW^2
    + q^2) \MW^2 \right) \mt^2 \MW^2 \non \\
  & \hspace{5em} + \left (2 \MTpq - 4 (\MW^2 + q^2) \MTps + 2
    \MW^4 + 7 \MW^2 q^2 \right) \MW^4 \Big ) A_0(\MW^2) \non \\
  & + \f{8 \sw^2 \left(2 \MTpq + (2 \MW^2 + q^2 ) \MTps - 4 \MW^4 -
      6 \MW^2 q^2 \right)}{\MW^2 q^2} B_0 (q^2,\MTps,\MTps) \non \\
  & - \f{8 \left (2 \sw^2 \mt^4 + (2 (\sw^2-3) \MW^2 +
      \sw^2 q^2) \hspace{0.5mm} \mt^2 - (2 \sw^2-3) (2 \MW^2 + 3 q^2) \MW^2
    \right)}{\MW^2 q^2} B_0(q^2,\mt^2,\mt^2) \non \\
  & - \f{24 \left (\MTps + \mt^2 - 2 \MW^2 - 3 q^2 \right)}{q^2}
  B_0(q^2,\mt^2,\MTps) \non \\
  & - \f{6 (\MTps - \mt^2) \left ( (4 \sw^2-2) (\MTps + \mt^2) + (4
      \sw^2 - 6) \MW^2 + (2 \sw^2 -1) q^2 \right)}{\MW^2 q^2}
  B_0(q^2,\MW^2,\MW^2) \\
  & - \f{16 \sw^2 \left (\MTpt + \MTpq q^2 - (3 \MW^2 + 2 q^2) \MTps
      \MW^2 + 2 (\MW^2 + q^2)^2 \MW^2 \right)}{\MW^2 q^2}
  C_0 (q^2,0,0,\MTps,\MTps,\MW^2) \non \\
  & + \f{8}{\MW^2 q^2} \Big (2 \sw^2 \mt^6 - (6 \MW^2 - (2 \sw^2-3)
  q^2) \hspace{0.5mm} \mt^4 - 2 (3 (\sw^2-2) \MW^2 + 2 (\sw^2-3)
  q^2) \hspace{0.5mm} \mt^2 \MW^2 \non \\
  & \hspace{5em} + 2 (2 \sw^2-3) (\MW^2 + q^2)^2
  \MW^2 \Big) C_0(q^2,0,0,\mt^2,\mt^2,\MW^2) \non \\
  & + \f{24}{\MW^2 q^2} \Big ( \left ( (2 \MW^2 + q^2) \MTps - 2
    (\MW^2 + q^2) \MW^2 \right) \hspace{0.5mm} \mt^2 \non \\
  & \hspace{5em} - 2 (\MW^2 + q^2) \left (\MTps - \MW^2 - q^2
  \right) \MW^2 \Big) C_0(q^2,0,0,\mt^2,\MTps,\MW^2) \non \\
  & - \f{12}{\MW^2 q^2} \Big ((2 \sw^2-1) \MTpt -\left (2 \MW^2 - (2
    \sw^2-1) q^2 \right) \MTpq - \left ((6 \sw^2-7)
    \MW^4 + 4 \sw^2 \MW^2 q^2 \right) \MTps \non \\
  & \hspace{5em} - 4 \, \cw^2 (\MW^2 + 2 q^2) \MW^4 \Big)
  C_0(q^2,0,0,\MW^2,\MW^2,\MTps) \non \\
  & + \f{12}{\MW^2 q^2} \Big ((2 \sw^2-1) \hspace{0.5mm} \mt^6 - \left
    (2 \MW^2 - (2 \sw^2-1) q^2 \right) \mt^4 - \left ((6
    \sw^2-7) \MW^4 + 4 \sw^2 \MW^2 q^2 \right) \mt^2 \non \\
  & \hspace{5em} - 4 \cw^2 (\MW^2+2 q^2) \MW^4 \Big )
  C_0(q^2,0,0,\MW^2,\MW^2,\mt^2) \scalebox{1.185}{\Bigg ]} \, , \non  
\end{align}
\end{widetext}
where $\MTp = f/v \, \mt/\sqrt{x_L (1 - x_L)}$ and $v \simeq 246 \,
\GeV $. Our new result for $C_{\rm LHT} (q^2)$ resembles for $q^2 = 0$
the analytic expression of the one-loop correction to the low-energy
$Z$-penguin function \cite{Blanke:2006eb}. Taking into account that
the latter result corresponds to unitary gauge while we work in 't
Hooft-Feynman gauge is crucial for this comparison.

\section{Numerical inputs}
\label{app:input}

In this appendix we collect the values of the experimental and
theoretical parameters used in our numerical analysis. The Higgs mass
and the various renormalization scales are scanned independently in
the ranges $100 \, \GeV < \mh < 600 \, \GeV$, $100 \, \GeV < \mut <
300 \, \GeV$, $40 \, \GeV < \muw < 160 \, \GeV$, $2.5 \, \GeV < \mub <
10 \, \GeV$, and $1 \, \GeV < \mu_c < 3 \, \GeV$, respectively.

The other parameters are displayed in
\Tabsand{tab:inputut}{tab:inputbrs}. Errors are indicated only if
varying a given parameter within its $1 \mysigma$ range causes an
effect larger than $\pm 0.1 \%$ on the corresponding result. When two 
errors are given, the first is treated as a Gaussian $1 \mysigma$ 
error and the second as a theoretical uncertainty that is scanned in
its range.    

\begin{table}[!t]
  \caption{\sf Parameters that enter the standard and universal
    UT analysis.}     
\vspace{-2.5mm} 
\begin{center}
\begin{tabular}{ccccc}
\hline \hline \\[-4.5mm]
Parameter & Value $\pm$ Error(s) & Reference \\[0.5mm] 
\hline 
$|V_{ud}|$ & $0.97377 \pm 0.00027$ & \cite{Yao:2006px} \\ 
$|V_{us}|$ & $0.2257 \pm 0.0021$ & \cite{Yao:2006px} \\ 
$|V_{cb}|$ & $(41.7 \pm  0.7) \times 10^{-3}$ & \cite{Yao:2006px} \\   
$|V_{ub}|$ & $(4.31 \pm 0.30) \times 10^{-3}$ & \cite{Yao:2006px} \\
$|\eps_K|$ & $(2.232 \pm 0.007) \times 10^{-3}$ & \cite{Yao:2006px} \\   
$\Delta m_K$ & $(3.4833 \pm 0.0059) \times 10^{-12} \, \MeV$ &
\cite{Yao:2006px} \\
$\Delta m_{B_d}$ & $(0.507 \pm 0.004) \, {\rm ps}^{-1}$ &
\cite{Barbiero:2007cr} \\  
$\Delta m_{B_s}$ & $(17.77 \pm 0.10 \pm 0.07) \, {\rm ps}^{-1}$
& \cite{unknown:2006ze} \\  
$\sin (2 \beta)_{b \to c \bar{c} s}$ & $0.675 \pm 0.026$ &
\cite{Barbiero:2007cr} \\ 
$m_{K^0}$ & $(497.648 \pm 0.022) \, \MeV$ & \cite{Yao:2006px} \\
$m_{B_d}$ & $(5.2793 \pm 0.0007) \, \GeV$ & \cite{Yao:2006px} \\
$m_{B_s}$ & $(5.3696 \pm 0.0024) \, \GeV$ & \cite{Yao:2006px} \\
$f_K$ & $(159.8 \pm 1.5) \, \MeV$ & \cite{Yao:2006px} \\
$f_{B_d} \hat{B}_{B_d}^{1/2}$ & $(244 \pm 26) \, \MeV$ &
\cite{Okamoto:2005zg} \\ 
$f_{B_s} \hat{B}_{B_s}^{1/2}$ & $(281 \pm 21) \, \MeV$ & 
\cite{Dalgic:2006gp} \\ 
$B_K$ & $0.79 \pm 0.04 \pm 0.09$ & \cite{Dawson:2005za} \\  
$\hat{B}_{B_d}$ & $1.28 \pm 0.04 \pm 0.09$ & \cite{Aoki:2003xb} \\
$\hat{B}_{B_s}$ & $1.30 \pm 0.03 \pm 0.09$ & \cite{Aoki:2003xb} \\
$\xi$ & $1.210^{+0.047}_{-0.035}$ & \cite{Okamoto:2005zg} \\
$\eta_{tt}$ & $0.5765 \pm 0.0065$ & \cite{Buras:1990fn, Herrlich} \\
$\eta_{ct}$ & $0.47 \pm 0.04$ & \cite{Herrlich} \\
$\eta_{cc}$ & $1.56 \pm 0.37$ & \cite{Herrlich} \\
$\eta_B$ & $0.551 \pm 0.007$ & \cite{Buras:1990fn} \\[0.5mm]
\hline \hline
\end{tabular}
\end{center}
\label{tab:inputut}
\end{table}

\Tab{tab:inputut} contains the quantities that are relevant for the
standard and universal UT analysis. We recall that there is a
discrepancy of around $1 \mysigma$ between the value of $|V_{ub}|$
obtained from inclusive and exclusive $\btouen$ transitions. Since
$|V_{ub}|$ only enters the standard UT analysis its actual value has
no impact on our main results. We therefore use the weighted average
of $|V_{ub}|$ given in \cite{Yao:2006px}. In the case of $f_{B_d}
\hat{B}_{B_d}^{1/2}$ and $\xi$ we take the values quoted in
\cite{Okamoto:2005zg}, which combines the values of the bag parameters
$\hat{B}_{B_d}$ determined by the JLQCD Collaboration using two light
flavors of improved Wilson quarks \cite{Aoki:2003xb} with the
staggered three flavor results for the $B$-meson decay constants
$f_{B_d}$ obtained by the HPQCD Collaboration \cite{Gray:2005ad}. The
central value and error of $f_{B_s} \hat{B}_{B_s}^{1/2}$ are taken from
the recent publication \cite{Dalgic:2006gp} of the HPQCD
Collaboration.  For a critical discussion of hadronic uncertainties in
the standard CKM fit we refer to \cite{Ball:2006xx}.

\begin{table}[!t]
\caption{\sf Quantities that are necessary for the calculation of the
  $\Ztobb$ POs and the rare and radiative $K$- and $B$-decays.}    
\vspace{-2.5mm} 
\begin{center}
\begin{tabular}{ccccc}
\hline \hline \\[-4.5mm]
Parameter & Value $\pm$ Error(s) & Reference \\[0.5mm] 
\hline 
$\GF$ & $1.16637 \times 10^{-5} \, \GeV^{-2}$ & \cite{Yao:2006px} \\
$\sws$ & $0.2324 \pm 0.0012$ & \cite{ewpm} \\
$\alpha_{\rm em} (0)$ & $1/137.036$ & \cite{Yao:2006px} \\ 
$\Daehad$ & $0.02768 \pm 0.00022$ & \cite{Hagiwara:2006jt} \\ 
$\as (\MZ)$ & $0.1189 \pm 0.0020$ & \cite{Yao:2006px, Bethke:2006ac} \\
$\MZ$ & $(91.1875 \pm 0.0021) \, \GeV$ & \cite{ewpm} \\
$\MW$ & $(80.405 \pm 0.030)\, \GeV$ & \cite{Yao:2006px} \\
$m_{t, {\rm pole}}$ & $(170.9 \pm 1.8) \, \GeV$ & \cite{Group:2007bx} \\
$m_b^{1 S}$ & $(4.68 \pm 0.03) \, \GeV$ & \cite{Bauer:2004ve} \\
$\mc (\mc)$ & $(1.224 \pm 0.017 \pm 0.054) \, \GeV$ &
\cite{Hoang:2005zw} \\ 
$m_\mu$ & $105.66 \, \MeV$ & \cite{Yao:2006px} \\[0.5mm]
$\BRXc$ & $0.1061 \pm 0.0017$ & \cite{Aubert:2004aw} \\
$C$ & $0.58 \pm 0.01$ & \cite{Bauer:2004ve} \\
$\lambda_1$ & $(-0.27 \pm 0.04) \, \GeV^2$ & \cite{Bauer:2004ve} \\
$\lambda_2$ & $0.12 \, \GeV^2$ & \cite{Yao:2006px} \\
$\kappa_L$ & $(2.229 \pm 0.017) \times 10^{-10}$ & \cite{Mescia:2007kn} \\
$\kappa_+$ & $(5.168 \pm 0.025) \times 10^{-11}$ & \cite{Mescia:2007kn}
\\
$\kappa_\mu$ & $(2.009 \pm 0.017) \times 10^{-9}$ &
\cite{Gorbahn:2006bm} \\ 
$\Delta_{\rm EM}$ & $-0.003$ & \cite{Mescia:2007kn} \\
$\tau(B_d)$ & $(1.527 \pm 0.008) \, {\rm ps}$ & \cite{Barbiero:2007cr}
\\ 
$\tau(B_s)$ & $(1.461 \pm 0.040) \, {\rm ps}$ & \cite{Barbiero:2007cr}
\\[0.5mm] 
\hline \hline
\end{tabular}
\end{center}
\label{tab:inputbrs}
\end{table}

\Tab{tab:inputbrs} summarizes the remaining parameters that enter the
determinations of the $\Ztobb$ POs and the rare and radiative $K$- and
$B$-decays. In the case of $\as (\MZ)$, we adopt the central value
from \cite{Bethke:2006ac}, but rescale the corresponding error by a
factor of two to be consistent with \cite{Yao:2006px}. We recall that
the parameters $\kappa_L$, $\kappa_+$, and $\kappa_\mu$ scale like
$(\lambda/0.225)^8$ and that the values given in \Tab{tab:inputbrs}
correspond to $\lambda \equiv |V_{us}| = 0.225$. This scaling has to
be taken into account in order to find consistent results in the case
of the rare $K$-decays. The IR finite long-distance QED correction
factor $\Delta_{\rm EM}$ entering the prediction of $\Kpnn$ accounts
for photon emission with energies of up to $20 \, \MeV$
\cite{Mescia:2007kn}. Since $\mc (\mc)$ and the phase-space factor $C
\equiv |V_{ub}/V_{cb}|^2 \, \Gamma(\BXuen)/\Gamma(\BXcen)$ are
strongly correlated we take both of their values from global analyses
of semileptonic $B$-decay spectra \cite{Bauer:2004ve,
  Hoang:2005zw}. However, to be conservative, we treat the errors of
$\mc (\mc)$ and $C$ as independent in our fit. If the anti-correlation
would be included, the individual uncertainties from $\mc (\mc)$ and
$C$ in the branching ratios of $\BXsga$, $\BXsll$, and $\BXdsnn$ would
cancel to a large extent against each other \cite{bsg}.

\end{document}